%% file: main.tex
\documentclass[amssymb,amsmath,prd,twocolumn,superscriptaddress,nofootinbib]{revtex4-2}

\usepackage{graphicx}
\usepackage{bm}
\usepackage{amssymb,amsmath}
\usepackage{mathrsfs}
\usepackage{latexsym}
\usepackage{color}
\usepackage[normalem]{ulem}
\usepackage{dcolumn}
\usepackage[colorlinks=true,citecolor=blue,urlcolor=blue]{hyperref}
\usepackage[usenames,dvipsnames]{xcolor}
\usepackage{xspace}
\usepackage{graphicx}
\usepackage{multirow}
\hypersetup{urlcolor=blue,citecolor=blue, linkcolor=blue,hypertexnames=false}

\input{macros}

\allowdisplaybreaks

\graphicspath{{./}{figures/}}

\newcommand{\CIT}{\affiliation{Department of Physics, California Institute of Technology, Pasadena, California 91125, USA}}
\newcommand{\CITLab}{\affiliation{LIGO Laboratory, California Institute of Technology, Pasadena, California 91125, USA}}
\newcommand{\Chicago}{\affiliation{Kavli Institute for Cosmological Physics, The University of Chicago, Chicago, Illinois 60637, USA}}
\newcommand{\UCBerkeley}{\affiliation{Department of Physics, University of California Berkeley, Berkeley, California 94720, USA}}

\definecolor{smgreen}{rgb}{0.26, 0.625, 0.277}

\newcommand{\chieff}{\chi_{\mathrm{eff}}}
\newcommand{\chip}{\chi_\mathrm{p}}

\newcommand{\sigmeas}{\sigma_\mathrm{meas}}
\newcommand{\Neff}{N_{\mathrm{eff}}}

\begin{document}


\title{Gravitational waves carry information beyond effective spin parameters but it is hard to extract}

\author{Simona J.~Miller}
\email{smiller@caltech.edu}
\CIT \CITLab
 
\author{Zoe Ko}  
\email{zko@berkeley.edu}
\UCBerkeley

\author{Tom Callister} 
\email{tcallister@uchicago.edu}
\Chicago

\author{Katerina Chatziioannou} 
\email{kchatziioannou@caltech.edu}
\CIT \CITLab

\date{\today}

\begin{abstract}
    Gravitational wave observations of binary black hole mergers probe their astrophysical origins via the binary spin, namely the spin magnitudes and directions of each component black hole, together described by six degrees of freedom. However, the emitted signals primarily depend on two effective spin parameters that condense the spin degrees of freedom to those parallel and those perpendicular to the orbital plane. Given this reduction in dimensionality between the physically relevant problem and what is typically measurable, we revisit the question of whether information about the component spin magnitudes and directions can successfully be recovered via gravitational-wave observations, or if we simply extrapolate information about the distributions of effective spin parameters.
    To this end, we simulate three astrophysical populations with the same underlying effective-spin distribution but different spin magnitude and tilt distributions, on which we conduct full individual-event and population-level parameter estimation. 
    We find that parameterized population models can indeed qualitatively distinguish between populations with different spin magnitude and tilt distributions at current sensitivity.
    However, it remains challenging to either accurately recover the \textit{true} distribution or to diagnose biases due to model misspecification.
    We attribute the former to practical challenges of dealing with high-dimensional posterior distributions, and the latter to the fact that each individual event carries very little information about the full six spin degrees of freedom.  
\end{abstract}

\maketitle


\section{Introduction}
\label{sec:intro}

The spins of black holes (BHs) in binaries (BBHs) are a unique probe of physics on multiple scales, from fundamental BH properties to stellar interiors and the astrophysical environments in which compact binaries form.
Each binary possesses six spin degrees of freedom: the spin magnitudes, polar angles (tilts) and azimuthal angles of each binary component~\cite{Blanchet:2013haa}. 
BH spins are encoded in the gravitational waves (GWs) the binary emits and can, at least in principle, be constrained from observation~\cite{O3a-pop,O3b-pop} by the LIGO~\cite{aLIGO} and Virgo~\cite{aVirgo} detectors. 
The magnitudes and directions of the spins at merger are determined by the spin each BH has upon formation as well as the binary's evolutionary history, e.g.~\cite{Gerosa:2017kvu,Rodriguez:2019huv,Bavera:2020inc}.
Spin measurements are therefore a promising way to determine whether BBHs form dynamically or in the field, e.g.~\citep{Vitale:2015tea,Rodriguez:2016vmx,Farr:2017uvj,Gerosa:2018wbw,Mandel:2018hfr,Zevin:2020gbd}, and answer questions such as the role of angular momentum transfer in stars, tidal interactions, and mass transfer, e.g.~\cite{Fuller:2019sxi,Qin:2018vaa,Bavera:2020uch,Steinle:2020xej,Zevin:2022wrw}. 

Despite their astrophysical importance, spins remain poorly constrained in GW data. Their imprint on the signal is typically subdominant to other intrinsic effects such as the BH masses, e.g.~\cite{Purrer:2015nkh,Vitale:2016avz,Chatziioannou:2018wqx,Biscoveanu:2021nvg,Miller:2023ncs}. Furthermore, not all six spin degrees of freedom affect the signal equally. 
Though waveform models formally depend on the full spin vectors~\cite{Khan:2019kot,Varma:2019csw,Pratten:2020ceb,Ossokine:2020kjp}, analytical post-Newtonian calculations indicate that the dominant spin effect is captured by two effective parameters: the effective aligned spin $\chieff$ that includes the spin components parallel to the Newtonian orbital angular momentum~\cite{Racine:2008qv}, and the effective precessing parameter $\chip$ that includes the perpendicular components~\cite{Schmidt:2014iyl}. 
The former primarily affects the length of the signal while the latter describes spin-precession, the change in binary orientation due to spin-orbit and spin-spin interactions~\cite{Apostolatos:1994}. 
Unsurprisingly, then, constraints on the astrophysical distributions of $\chieff$ and $\chip$ can be typically obtained with fewer observations and are less prone to population model systematics than the spin components~\cite{Miller:2020zox,Roulet:2021hcu,Callister:2022qwb}.

Although less well measurable, it is instead the underlying spin components that are of prime astrophysical interest. 
GW signals contain \textit{some} information about component spins.
However, unlike $\chieff$ and $\chip$ that appear prominently in the GW phase and amplitude and whose measurability can be predicted with analytic arguments~\cite{Ng:2018neg,Chatziioannou:2020msi}, individual spin components have a significantly subdominant effect on the waveform.
The resulting constraints on the \textit{astrophysical} distribution of spin components are correspondingly weaker and in many cases subject to uncertainties about the role of population models~\cite{Callister:2022qwb,Vitale:2022dpa}. 
Indeed, even though it is widely accepted that BBHs have a range of $\chieff$ values that are not symmetric about zero~\cite{O2-pop,O3a-pop,O3b-pop} and that not all BBHs have a vanishing $\chip$~\cite{O3a-pop,O3b-pop}, the exact shape of the inferred distribution for spin magnitudes and directions depends on the parameterization of the corresponding population model. 
For example, different parametrizations for the angle between the spins and the Newtonian orbital angular momentum lead to varied conclusions about where the distribution peaks and the degree of spin-orbit misalignment~\cite{O3b-pop,Callister:2022qwb,Golomb:2022bon,Vitale:2022dpa,Tong:2022iws,Edelman:2022ydv}.

Central to this discussion are the questions of how much information GW signals actually contain about the BH spin components versus $\chieff$ alone, how feasible it practically is to reliably extract this information, and the extent to which conclusions are driven by informative data or simply by overly restrictive models.
In this paper, we approach these issues by posing three questions, from which we conclude:
\begin{enumerate}
    \item \textbf{Do GWs carry information about spin components, or are we just extrapolating the effective aligned spin $\chieff$? }(Sec.~\ref{sec:measuring})
    
    \emph{Yes, we can distinguish between populations with low, moderate, and high spins even when they have identical effective spin distributions.}
    \item  \textbf{Can component spin distributions be \textit{accurately} measured?} (Sec.~\ref{sec:pe_bias})

    \emph{Even though we can qualitatively tell apart BH populations with different spin distributions, characterizing them accurately is practically challenging.}

    \item \textbf{Can we tell when measurements of component spin distributions are biased?} (Sec.~\ref{sec:ppcs})

    \emph{Common tests based on posterior predictive checks cannot identify modeling biases in component spin distributions due to the fact that individual-event posteriors are extremely weakly informative about spin components.}
    
\end{enumerate}

The remainder of this paper presents our analysis in support of these conclusions. 
We discuss spin degrees of freedom, effective spin parameters, and the notation used throughout in Sec.~\ref{sec:spins}. 
Our Methods are briefly described in Sec.~\ref{sec:methods}, and are expanded upon in the Appendices.
Results about measuring the component spin distributions are presented in Sec.~\ref{sec:measuring}.
Section~\ref{sec:pe_bias} introduces the extensive series of verification methods -- both population and individual-event level -- we use to ensure the robustness of our results, all of which are further elaborated upon in the Appendices.  
In Sec.~\ref{sec:ppcs}, we identify limitations of the traditional method of using posterior predictive checks to assess biased population measurements, and identify the sources of this bias.
We compare our findings to those of past work in Sec.~\ref{sec:discussion}, and then conclude in Sec.~\ref{sec:conclusions}.

\section{Spin magnitudes and tilts versus effective spin parameters}
\label{sec:spins}

Each BH in the binary is described by a dimensionless spin vector $\vec \chi_i$, $i\in\{1,2\}$.
In a coordinate system where the $z$-axis is aligned with the binary Newtonian orbital angular momentum $\vec L$, the spin vector is characterized by a magnitude $\chi_i \in [0,1]$, polar angle $\theta_i \in [0, \pi]$, and azimuthal angle $\phi_i \in [0,2\pi]$. Modulo horizon absorption effects, the spin magnitude is constant throughout the binary evolution~\cite{Alvi:2001mx,Poisson:2004cw}, while the spin angles evolve due to spin-orbit and spin-spin interactions causing the spin vector to precess~\cite{Apostolatos:1994,Gerosa:2015tea}.

The full six spin degrees of freedom remain relatively poorly constrained by GW signals. Rather, the dominant spin effects are expressed by two effective parameters. The mass-weighted average spin projected onto $\vec L$
\begin{equation}\label{eqn:chieff}
    \chieff = \frac{\chi_1 \cos\theta_1 + q\chi_2 \cos\theta_2}{1+q}\in (-1,1)\,, 
\end{equation}
is referred to as \emph{effective aligned spin}, where we have defined the binary mass ratio $q \equiv m_2/m_1$, where $m_1 \geq m_2$ are the BH masses.
The effective aligned spin is, in general, better constrained as it is related (in the equal mass limit) to the leading-order spin contribution in the post-Newtonian expansion for the GW inspiral phase~\cite{Blanchet:2013haa}. Additionally, $\chieff$ is conserved under spin-precession and radiation reaction to at least the second post-Newtonian order~\cite{Racine:2008qv}. 

Spin-precession effects are captured with the \emph{effective precessing parameter}
\begin{equation}\label{eqn:chip}
    \chi_\mathrm{p} = \mathrm{max}\left[\chi_1 \sin\theta_1,
    \left(\frac{3+4q}{4+3q}\right) q \,\chi_2 \sin\theta_2\right]\in [0,1) \,\,.
\end{equation}
This parameter and its extensions~\cite{Gerosa:2020aiw,Thomas:2020uqj} are motivated by the fact that spin-orbit precession (and the GW amplitude and phase modulations it induces) are driven by in-plane spin components~\cite{Schmidt:2012rh,Schmidt:2014iyl}. Constraints on $\chi_\mathrm{p}$ are typically much weaker than $\chieff$ especially given the observed absence of large spin-precession in BBHs~\cite{O3a-pop,O3b-pop}. In what follows, we therefore focus on $\chieff$.

\section{Methodology}
\label{sec:methods}

%
\begin{figure*}
    \centering
    \includegraphics[width=\textwidth]{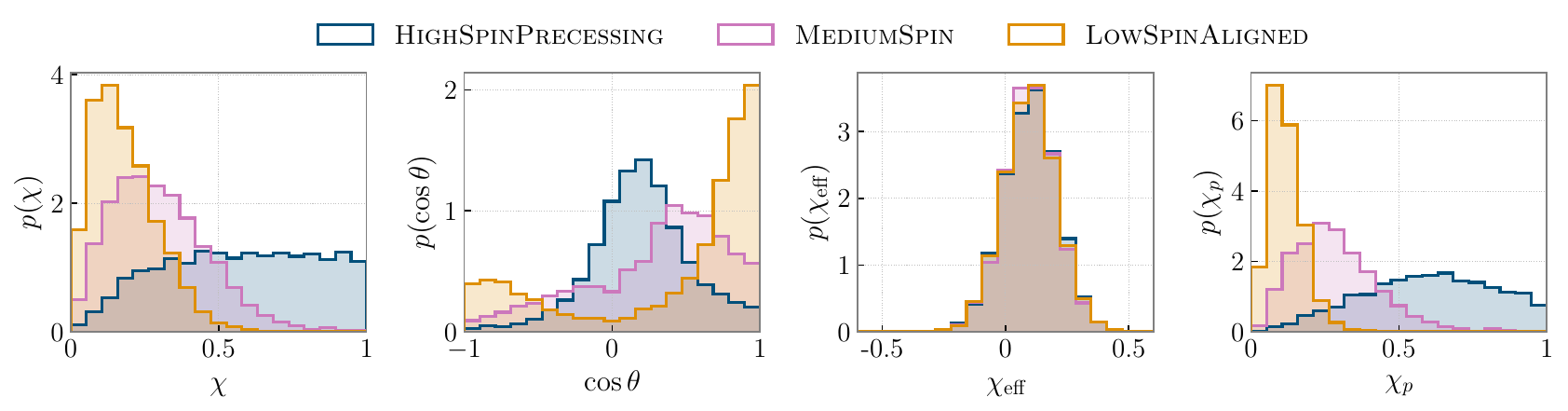}
    \caption{Spin distributions for the three simulated BBH populations we use to assess the amount of recoverable information GW signals contain about component spin distributions. The three populations share the same $\chieff$ distribution (chosen for consistency with current data), but differ in their underlying component spin and $\chip$ distributions. From left to right, panels show the spin magnitude $\chi$, spin tilt $\cos\theta$, the effective spin parameter $\chieff$, and the effective precession parameter $\chip$. (\textit{Navy}) \textsc{HighSpinPrecessing}: a BBH population with few vanishing spins that are preferentially oriented close to the orbital plane.  (\textit{Pink}) \textsc{MediumSpin}: a BBH population with moderate spin magnitudes peaking at $\chi=\popTwoChiPeak$ and preferentially aligned with the Newtonian orbital angular momentum.  (\textit{Orange}) \textsc{LowSpinAligned}: a BBH population with low spin magnitudes peaking at $\chi=\popThreeChiPeak$ with both strongly aligned and anti-aligned sub-populations.}
    \label{fig:populations}
\end{figure*}

In order to isolate the amount of information included in GW signals about the spin components relative to the effective spin, we simulate astrophysical populations with identical $\chieff$ distributions but different underlying component spin distributions.
We choose a $\chieff$ distribution that is qualitatively similar to current constraints~\cite{O3b-pop, Callister:2022qwb, Callister:2023tgi} and decompose it into three populations with distinct spin magnitudes and tilt angle distributions. 
The azimuthal angles are uniformly distributed.
These distributions are not astrophysically motivated, but rather selected as distinct test cases of potential distributions.

The three simulated astrophysical distributions are shown in Fig.~\ref{fig:populations}, with further details given in Appendix~\ref{appendix:pops}: 
\begin{itemize}
    \item The \textsc{HighSpinPrecessing} population contains BHs with the most extremal spins and tilts: the majority of the population has $\chi > 0.5$ and tilts nearly in-plane, corresponding to significant spin-precession. 
    \item The \textsc{MediumSpin} population is most similar to current constraints: preferentially small to moderate spin magnitudes peaking at $\popTwoChiPeak$, and a wide range of tilts with a preference for alignment compared to anti-alignment.
    \item  The \textsc{LowSpinAligned} population has the smallest spin magnitudes, with nearly all BHs having $\chi < 0.5$. 
    Uniquely, this population has a bimodal spin tilt angle distribution, with a larger peak at $\cos \theta = 1$ (perfect alignment) and a smaller peak at $\cos\theta = -1$ (perfect anti-alignment). 
    It is therefore a test case of sensitivity to mixture models. 
\end{itemize}

With these three populations, we conduct a full end-to-end injection/recovery campaign. 
We draw parameters describing individual GW events from each distribution, restrict to detectable events with a network optimal signal-to-noise ratio (SNR) above $10$ in the LIGO Livingston, LIGO Hanford, and Virgo detectors, simulate data assuming O3 sensitivity~\cite{O3_PSDs}, and obtain samples from the multidimensional posterior distribution of the binary parameters for each event individually. 
We then hierarchically model the population distribution of the simulated posteriors with parametrized population models.

The individual-event posterior sampling is conducted with the nested sampler \textsc{Dynesty}~\cite{dynesty} as implemented in \textsc{Bilby}~\cite{bilby, Romero-Shaw:2020owr}.
We use the \textsc{IMRPhenomXPHM} waveform model~\cite{Pratten:2020ceb} both for simulation and recovery as it models all six spin degrees of freedom, contains higher order radiation modes, and is the least computationally expensive option available.
Although more computationally expensive than approximate parameter estimation~\cite{Fishbach:2019ckx,Fairhurst:2023idl,Farah:2023vsc}, it is essential to use full stochastic sampling for this work.
As we are trying to discern subtle effects in the signals, we must properly characterize the individual-event likelihoods.
Full details about parameter estimation settings are given in Appendix~\ref{appendix:pe}.
For hierarchical inference, we primarily use the Markov Chain Monte Carlo sampler \textsc{Emcee}~\cite{emcee}, with some follow-up studies run with \textsc{Numpyro}~\cite{numpyro1,numpyro2}.
The full hierarchical inference procedure is outlined in Appendix~\ref{appendix:inference}, with the parameterized population models detailed in Appendix~\ref{appendix:models}.

For simplicity, the hierarchical inference ignores the azimuthal angles and in what follows use the term ``spin components" to refer to the spin magnitudes and tilt angles.
The parameter estimation prior and the population distribution for
the azimuthal angles coincide, therefore fixing their distribution (to truth) does not incur a bias, more details are available in Appendix~\ref{appendix:models}.

\section{Different spin magnitude and tilt distributions can be distinguished}
\label{sec:measuring}

%
\begin{figure*}[t]
    \centering
    \includegraphics[width=\textwidth]{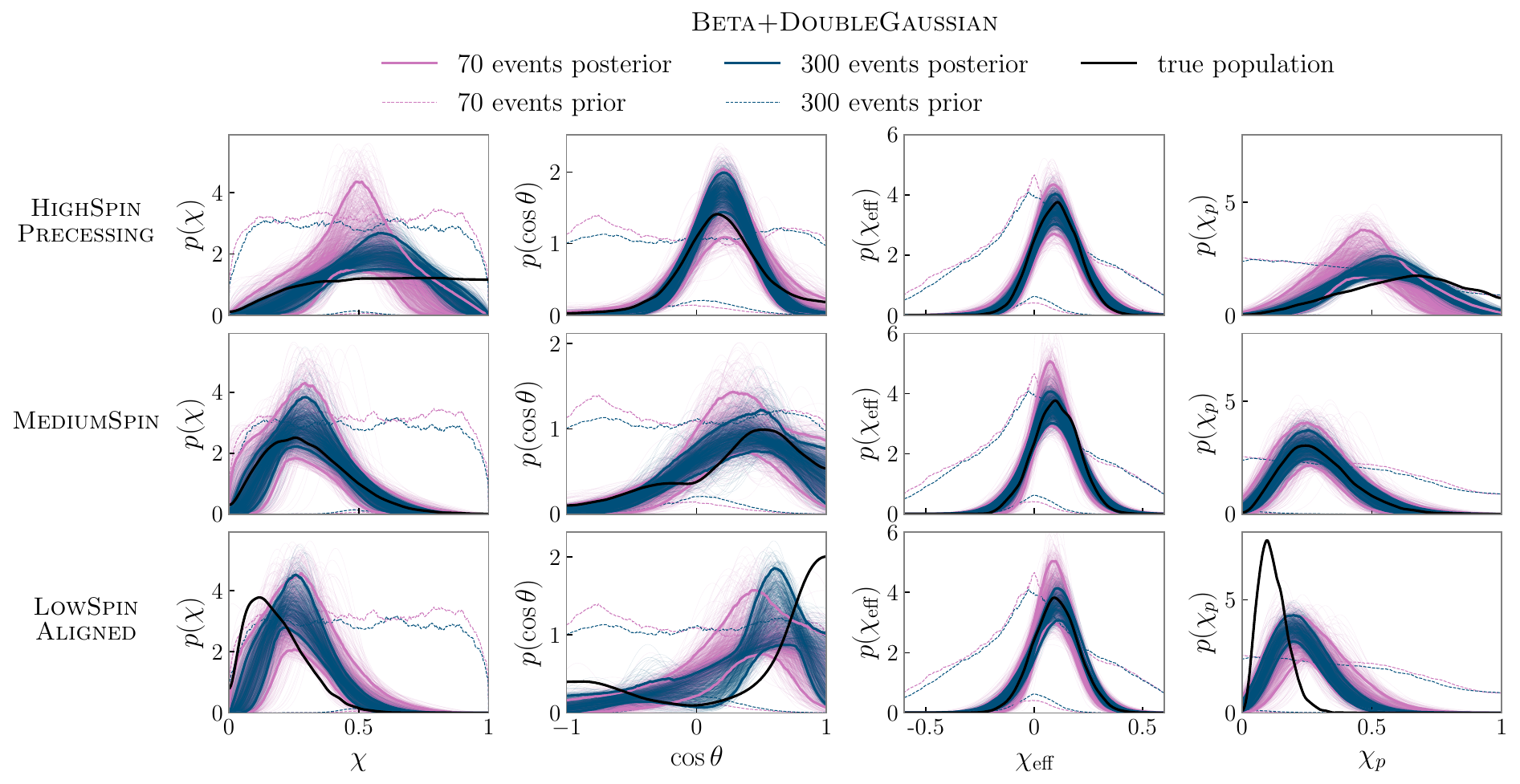}
    \caption{
    Inferred distributions for various spin parameters (from left to right: spin magnitude $\chi$, spin tilt $\cos\theta$, effective $\chieff$ spin, effective precessing spin $\chip$) and the three simulated populations (top to bottom). All results are obtained with the \textsc{Beta+DoubleGaussian} model using 70 (pink) and 300 (navy) events. Traces correspond to draws from the population posterior and solid lines enclose 90\% of the probability. The black solid line corresponds to the true, underlying population.
    The dashed lines show the 90\% credible intervals inferred by sampling the prior on the population-level parameters, including the effective sample cut as defined in Eq.~\eqref{eq:Neff}.
    }
\label{fig:betaPlusDoubleGaussianTrace}
\end{figure*}

Using the results of the signal injection and parameter estimation campaign, we perform hierarchical inference~\cite{Loredo:2004nn,Mandel:2018mve,Vitale:2020aaz} on events drawn from each population shown in Fig.~\ref{fig:populations} in order to reconstruct their underlying spin distributions. 
Population inference requires the adoption of a model for the component spin and tilt distributions. 
We select an analytic model in which spin magnitudes follow a non-singular Beta distribution and the spin tilt angles follow a bimodal Gaussian distribution, which we hereby refer to as the \textsc{Beta+DoubleGaussian}; see Appendix~\ref{appendix:betaPlusDoubleGaussian} for a full description.
We choose this population model for its simplicity and similarity to common models in the literature, albeit with small modifications to target our questions of interest.
While the true, underlying distributions shown in Fig.~\ref{fig:populations} were \textit{not} explicitly drawn from the \textsc{Beta+DoubleGaussian} model, this model is expected to agree with each simulated population to within statistical uncertainties.\footnote{We confirm that the \textsc{Beta+DoubleGaussian} model is a good fit to the underlying populations through a least-squares fit. 
The Kullback-Leibler divergences~\cite{Kullback:1951} between the best fit and true underlying distributions are $<10^{-4}$ for all spin magnitude distributions and $<0.08$ for all cosine tilt distributions.
As a further check, we also run hierarchical inference on catalogs of simulated, Gaussian individual-event spin posteriors, with results discussed in Appendix~\ref{appendix:mock_posteriors}.}
We do not model the spin azimuthal angle, effectively (and correctly) assuming that it is distributed according to its uniform prior.

Fig.~\ref{fig:betaPlusDoubleGaussianTrace} shows the inferred distributions for various spin parameters for the three simulated populations under the \textsc{Beta+DoubleGaussian} population model. 
The results from a 70-event catalog are plotted in pink, and from a 300-event catalog in navy, chosen to mimic O3 and projected O4 catalog size respectively.
Black, bold traces show the true underlying populations for comparison. 
The $\chi$ (first column) and $\cos\theta$ (second column) distributions are generated by random draws from the posteriors on the population parameters, shown in Figs.~\ref{fig:cornerplots1}, \ref{fig:cornerplots2}, and \ref{fig:cornerplots3} in Appendix~\ref{appendix:cornerplots}. 

The $\chieff$ (third column) and $\chip$ (fourth column) distributions are generated by \textit{(i)} randomly drawing from the inferred $\chi$, $\cos\theta$, and $q$ distributions, \textit{(ii)} calculating the effective spin parameters from these draws, and \textit{(iii)} generating a Gaussian kernel density estimate.
We find that \textit{the $\chieff$ distributions are reconstructed accurately across all three populations and for both catalog sizes.}

Switching to the component spins, we can qualitatively distinguish between their distributions for each population. 
The spin magnitude inferred for the \textsc{HighSpinPrecessing} population is the widest and has the largest mean. 
From~\textsc{HighSpinPrecessing}~to~\textsc{MediumSpin}~to~\textsc{LowSpinAligned}, the means and widths of the inferred distributions get progressively smaller, as is the case for the true, underlying populations. 
The mass ratio distributions are also successfully recovered for all three populations and two catalog sizes, as shown in  Appendix~\ref{appendix:cornerplots}. 
We, therefore, \emph{can distinguish between populations with low, moderate, and high spins when they have identical $\chieff$ distributions}.

Although we can qualitatively characterize the spin magnitude and tilt distributions among these three populations, in some cases we cannot reliably characterize their properties accurately. 
Specifically, the true underlying distribution of the \textsc{LowSpinAligned} population does not lie within the 90\% credible reconstructed region as measured with the  \textsc{Beta+DoubleGaussian} model. 
The bimodality of this population's tilt angle distribution is not recovered and the inferred spin magnitude distribution has a higher mean than truth. 
We confirm that this mismatch between the true and inferred distributions is not \textit{solely} driven by overly restrictive priors on the population-level parameters, plotted with dashed lines in Fig.~\ref{fig:betaPlusDoubleGaussianTrace}.
For example, the bias in the inferred spin magnitude distribution for the \textsc{HighSpinPrecessing} population occurs over a region where the prior generates a flat distribution. 

Population measurements of the effective spin are more robust against bias than component spins. 
Even a notable mismatch between the true and recovered spin magnitude and tilt distributions for the \textsc{LowSpinAligned} population results in a precisely and accurately constrained $\chieff$ distribution.
The $\chip$ distributions are more susceptible to inaccurate recovery in the component spins, effectively inheriting their biases. 
For example, for the $\chip$ distribution of the \textsc{LowSpinAligned} population: the inference of more in-plane spin ($\cos\theta \sim 0$) combined with the slight over-estimation of the mean of the spin magnitude distribution, leads to a corresponding over-estimate of the bulk of the $\chip$ distribution.
This means that $\chieff$, but not necessarily $\chip$ can be reliably characterized on a population level by component spin measurements.
We additionally look at alternative definitions of $\chip$ (see e.g.~\citet{Gerosa:2020aiw}) and obtain over-all consistent results with the standard $\chip$ given in Eq.~\ref{eqn:chip}.

\section{Difficulties of Measuring Component Spin Distributions}
\label{sec:pe_bias}

The biased reconstruction of the \textsc{LowSpinAligned} population is unexpected. 
The injection and recovery campaign was performed using the same waveform model, and selection effects were self-consistently handled in both signal selection and parameter estimation. 
Under these conditions, there is no \textit{a priori} reason why population recovery should fail.
As such, our results instead suggest a shortcoming in either the parameter estimation or population recovery stages of the analysis. 

To diagnose this shortcoming, we employ a slew of checks, all of which are further elaborated upon in Appendix~\ref{appendix:verification}.
To ensure that the problem is not our hierarchical inference framework and implementation we do the following: 
\begin{itemize}
    \item \textbf{Simulate Gaussian individual-event spin posteriors} (Appendix~\ref{appendix:mock_posteriors}) -- For each of the 300 events per population, we generate a series of simulated Gaussian individual-event spin posteriors with a range of measurement errors and underlying correlations, and use these as input to hierarchical inference. In these cases, the \textsc{Beta+DoubleGaussian} population model \textit{is} able to recover the underlying populations, as seen in Fig.~\ref{fig:mockPopsBetaDoubleGaussian}. 
    
    \textit{Implications}: The hierarchical inference and selection effect framework is algorithmically robust, and the \textsc{Beta+DoubleGaussian} model is able to recover the true population distributions to within statistical uncertainties. 
    \item \textbf{Fix either the spin magnitude or tilt angle distribution to the truth} (Appendix~\ref{appendix:reducing_complexity}) -- When only fitting for the $\chi$ or $\cos\theta$ population and not the other, we are still unable to recover the correct distribution for the \textsc{LowSpinAligned} population; see Fig.~\ref{fig:fixed_chi_or_cost_dist}.
    
    \textit{Implications}: The observed bias in the spin magnitude and tilt angle distributions is not related to correlations between the two distributions.
    \item \textbf{Use a different sampler for the hierarchical likelihood} (Appendix~\ref{appendix:alt-codes}) -- 
    We repeat the analysis of Fig.~\ref{fig:betaPlusDoubleGaussianTrace} with an independently-implemented hierarchical inference code that is based on \textsc{Numpyro} instead of \textsc{emcee}. We obtain essentially identical results, shown in Fig.~\ref{fig:spins-with-masses}.
    
    \textit{Implications}: The hierarchical inference and selection effect framework is algorithmically robust.
     \item \textbf{Fit for the mass and redshift distributions instead of fixing it to truth} (Appendix~\ref{appendix:masses}) -- Our main results fit for the spin magnitude, spin tilt, and mass ratio distributions, while fixing the distributions of the primary mass and redshift to their true population values, given in Appendix~\ref{appendix:pops}. Figure~\ref{fig:spins-with-masses} extends these results to also fit for the mass and redshift distributions and shows the corresponding spin population posteriors, which remain unchanged. The mass and redshift distributions are recovered with no bias.
    
    \textit{Implications}: We have not misspecified the mass or redshift distributions when fixing them to truth during hierarchical inference, nor biased results of our spin inference by neglecting to simultaneously fit for the mass and redshift distributions.
    \item \textbf{Plot rates instead of probability distributions} (Appendix~\ref{appendix:rates}) -- 
    Apparent disagreement between injected and recovered probability distributions can sometimes be caused by comparing injected and recovered probability distributions, rather than differential merger rate densities.
    In the main text we do not infer the overall rates of black hole mergers, but only the shapes of their spin distributions.
    To check if neglecting the merger rate contributes to apparent disagreement between injected and recovered populations, we repeat our hierarchical inference while also fitting for the rate of black hole mergers as function of spin magnitude and tilt.
    This yields the results shown in Fig.~\ref{fig:spins-with-rates}, which remain qualitatively similar to those in Fig.~\ref{fig:betaPlusDoubleGaussianTrace}.

    \textit{Implications}:
    We have not biased results of our spin measurements by failing to measure or plot absolute merger rates, rather than probability distributions.
    \item \textbf{Exclude spin selection effects} (Appendix~\ref{appendix:misc}) -- The selection function only negligibly affects spin magnitudes and tilt angles. To ensure that we are not incorrectly implementing the selection function in the hierarchical likelihood, we conduct hierarchical inference without including selection effects in spin. This does not impact our results, as can be seen in Fig.~\ref{fig:misc_tests}.

    \textit{Implications}: The implementation of the selection function is algorithmically robust.
    \item \textbf{Employ different methods of breaking the degeneracy in the bimodal Gaussian model} (Appendix~\ref{appendix:misc}) -- For a bimodal distribution, some method must be imposed to break the degeneracy between the two components of the model. For the \textsc{Beta+DoubleGaussian} model, this can be done in one of three ways: imposing an ordering of the means, the widths, or limiting the mixing fraction be $\leq0.5$. Sometimes one method of breaking the degeneracy converges better than another. As shown in Fig.~\ref{fig:misc_tests}, we find that this is not the case here and different methods perform comparably. 

    \textit{Implications}: Our choice of degeneracy-breaking between the two Gaussian components in our population model is not causing convergence issues. 
    \item \textbf{Run hierarchical inference on different 70-event catalog instantiations} (Appendix~\ref{appendix:misc}) -- Finally, to get a sense of how much the specific 70 events we select from the underlying population affect hierarchical inference, we repeat the procedure with several different catalog instantiations. While there is expected variance in the results --see Fig.~\ref{fig:diff_catalog_instantiations}-- it cannot account for the degree of mismatch seen in the bottom row of Fig.~\ref{fig:betaPlusDoubleGaussianTrace}. Additionally, each catalog instantiation leads to a different number of per-event effective samples, which we find are not correlated to the goodness of fit.

    \textit{Implications}: The observed bias does not arise from an insufficient number of per-event effective samples. 

\end{itemize}
We then move towards investigating the underlying individual-event parameter estimation with the following checks: 
\begin{itemize}
    \item \textbf{Sampler settings} -- We run parameter estimation with a large variety of sampler settings in \textsc{Bilby}, and eventually adopt the standard, reviewed settings for our headline results of Fig.~\ref{fig:betaPlusDoubleGaussianTrace}. 

    \textit{Implications}: Running with more aggressive sampler settings in \textsc{Bilby} may fix convergence problems, but this was not the case for any configurations we employed. 
    \item \textbf{Probability-probability plots} (Appendix~\ref{appendix:PPplots}) -- We generate probability-probability (P-P) plots~\cite{Cook:2006,Talts:2018} for reweighted individual-event \textsc{Bilby} posteriors. 
    As seen in Fig.~\ref{fig:PPplots}, the test passes. 
    
    \textit{Implications}: Either the \textsc{Bilby} individual-event posterior samples are unbiased, or the biases are subtle enough to not be detectable by a P-P test, as warned against in~\cite{Biscoveanu:2021eht}.
    \item \textbf{Use a different waveform model} (Appendix~\ref{appendix:reducing_complexity}) -- We re-run individual-event inference on the same sets of events with \textsc{Bilby} using the \textsc{IMRPhenomXP} waveform model instead of \textsc{IMRPhenomXPHM} both for injection and recovery. Results with this waveform model are comparable or worse to that presented in the main text with \textsc{IMRPhenomXPHM}, although the bimodality of the \textsc{LowSpinAligned} population is slightly better constrained; see Fig.~\ref{fig:4D_and_8D_and_XP}.

    \textit{Implications}: The \textit{existence} of bias in the measured spin magnitude and tilt distributions is not driven by our choice of waveform model, although the specific details of how that bias manifests appear to be, i.e.~different waveforms yield different population-level results. This indicates that the bias may be due to individual-event sampling issues.
    \item \textbf{Fix non-spin parameters to truth in individual-event sampling} (Appendix~\ref{appendix:reducing_complexity}) -- Finally, we conduct individual-event inference with \textsc{IMRPhenomXPHM} fixing all parameters aside from the spin magnitudes to tilt angles to truth (i.e.~use delta function priors at their injected values). In this case, the \textsc{Beta+DoubleGaussian} population model \textit{is} able to successfully recover the truth for all three populations; see Fig.~\ref{fig:4D_and_8D_and_XP}.

    \textit{Implications}: The added complexity going from sampling just spins to all fifteen binary parameters is a likely culprit for the biased spin magnitude and tilt angle distributions. 
\end{itemize}

Although we can qualitatively tell apart the different populations in Fig.~\ref{fig:populations}, our results indicate that the spin distribution of all possible BBH populations cannot necessarily be \textit{accurately} measured under the range of analyses considered in this work.
Full parameter estimation with spin-precession is a technically challenging analysis. 
Despite conducting tens of model checking procedures, we cannot fully identify the driving source of the bias observed in Fig.~\ref{fig:betaPlusDoubleGaussianTrace}. 
We hypothesize that the error is due to issues related to sampling from the high-dimensional posterior for individual events, as suggested by the final bulletpoint above. 
If the issue with unbiased recovery is indeed due to poor convergence of parameter estimation, then it is possible that future algorithmic improvements in parameter estimation will resolve things and allow for accurate recovery.

\section{Identifying bias is difficult: limitations of posterior predictive checks}
\label{sec:ppcs}

\begin{figure*}
    \centering
    \includegraphics[width=\linewidth]{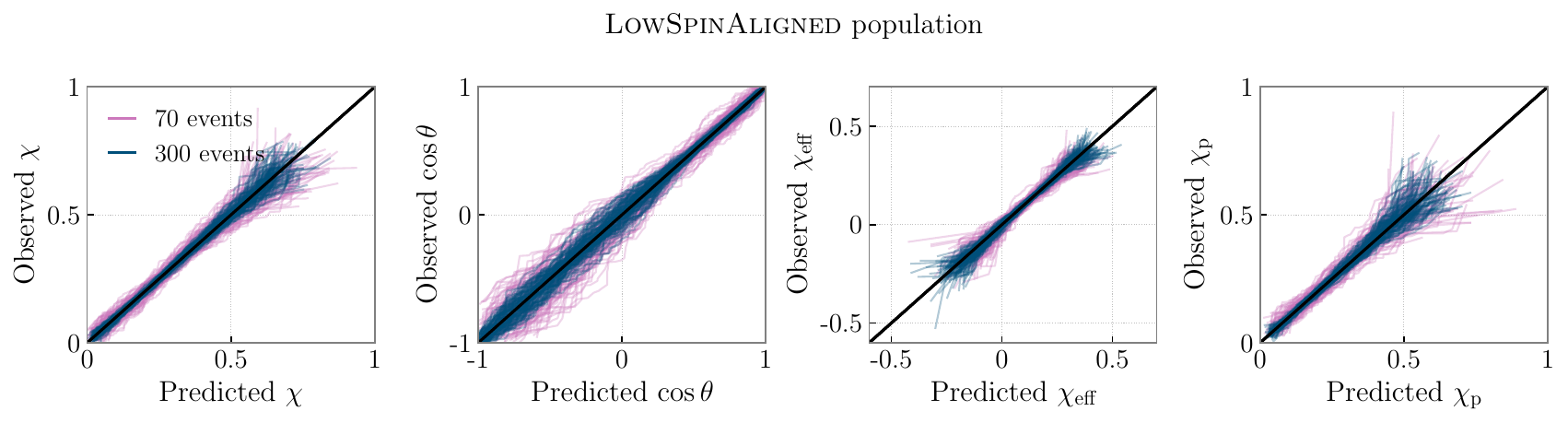}
    \caption{
     Posterior predictive checks (PCCs) for spin parameters (left to right: component spin magnitudes $\chi$, component spin tilts $\cos\theta$, effective spin $\chieff$, and effective precessing spin $\chip$) of the \textsc{LowSpinAligned} population under the \textsc{Beta+DoubleGaussian} population model. 
     Each trace is one catalog from the observed and the predicted populations. 
     100 catalogs are shown, for results with 70 (pink) and 300 (navy) events. 
     In the absence of discrepancy between the inferred and true population distributions, the traces should on average follow the diagonal.
     However, here there \textit{is} such a discrepancy, as can be seen in the bottom row of Fig.~\ref{fig:betaPlusDoubleGaussianTrace}, and the traces do on average follow the diagonal, meaning that this PPC is \textit{not necessarily a good diagnostic tool} for component spins.
    }
    \label{fig:PPCs}
\end{figure*}

\begin{figure*}
\centering
    \includegraphics[width=\linewidth]{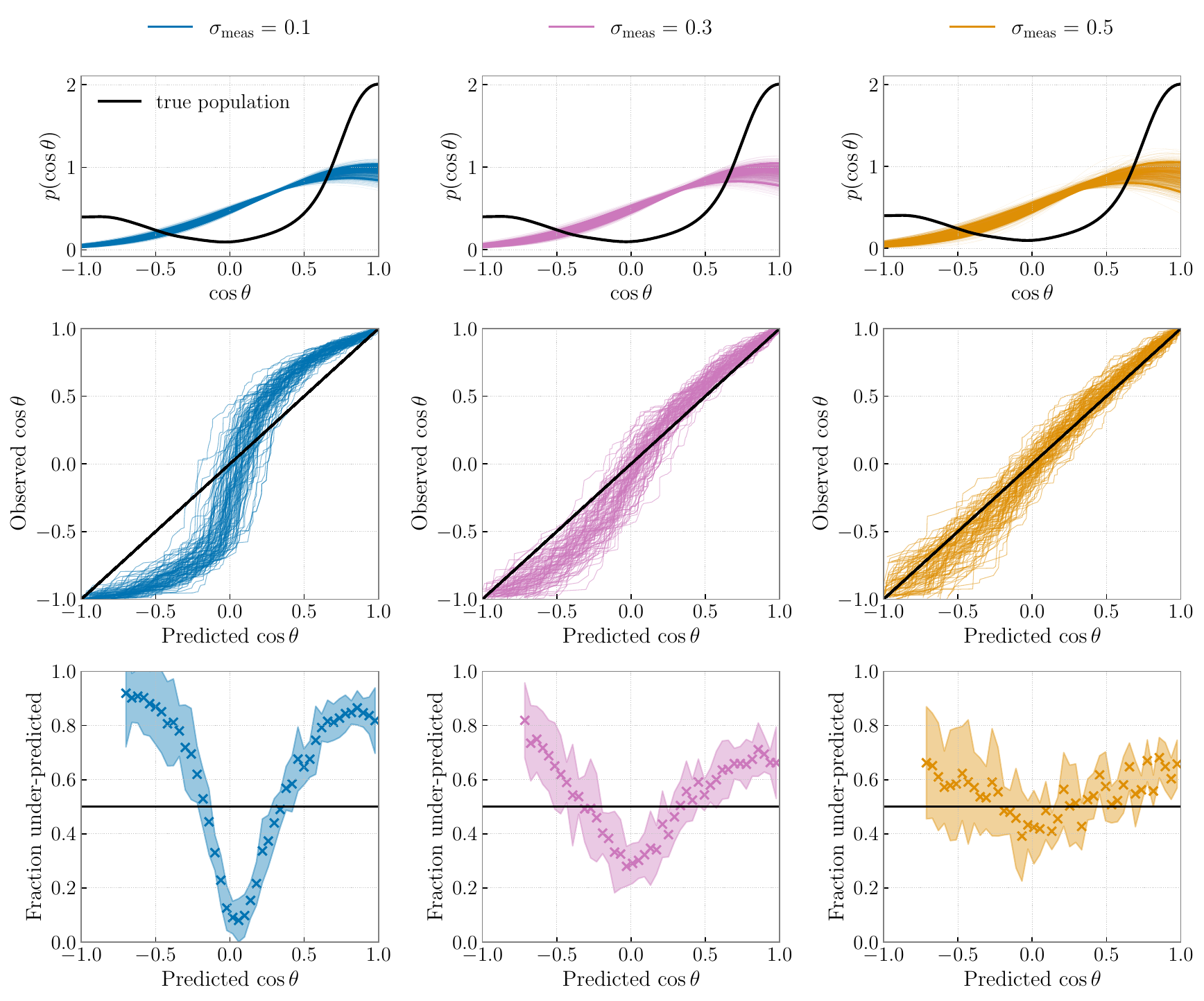}
    \caption{
    Results for diagnosing model misspecification for the \textsc{LowSpinAligned} population's $\cos\theta$ distribution under the \textsc{Beta+Gaussian} model.
    Results from three different simulated measurement uncertainties are shown: individual-event spin measurement error of $\sigmeas =$ 0.1 (blue), 0.3 (pink), and 0.5 (yellow).
    The ``true" measurement error from the \textsc{Bilby} runs averages to $\sigmeas=\trueSigmaMeas$.
    All results are shown from runs done on 70-event catalogs.
    \textit{(Top Row)}
    Traces corresponding to draws from the population posterior, compared to the true population (black).
    \textit{(Middle Row)}
     PPCs from 100 catalogs, each with 70 events.
     In the absence of model misspecification, the traces should on average follow the diagonal.
     \textit{(Bottom Row)}
     Fraction of events from the posterior predictive checks with $\cos\theta$ under-predicted.
     The shaded regions indicate three-sigma uncertainty on each average value, marked by the crosses.
     In the absence of model misspecification, the error bars should encompass a horizontal line at 0.50.
    }
    \label{fig:mockPopsPPCs}
\end{figure*}

While the \textsc{Beta+DoubleGaussian} model was able to produce qualitatively correct results for each of the three distinct populations, it was not able to successfully recover the true underlying populations.
This leads us to the question: \textit{can commonly-used modeling diagnostics successfully identify poorly performing fits to component spin distributions?}
There are multiple avenues through which a population model can fail: either the model is theoretically a good fit and for any number of reasons (e.g.~those discussed in Sec.~\ref{sec:pe_bias}) cannot find the truth, or the model is intrinsically a poor fit, i.e.~it does not have enough flexibility to find the shape of the true, underlying distribution.
Both cases induce mismatch between the true distribution and the inferred distribution, which we hope to diagnose using only the information available to us.
In this section, we begin by discussing the first scenario (Fig.~\ref{fig:PPCs}) and then the second (Fig.~\ref{fig:mockPopsPPCs}).

In reality, given the complexities of astrophysical BH spin evolution, it is almost certain that our measured distributions are in some other way discrepant with the truth; phenomenological models likely cannot perfectly reflect the underlying populations.
Model checks on current data sets are then used to motivate more complicated parametric models that do not suffer from identifiable deficiencies. 
In parallel, nonparametric inference introduces more flexible models that are based on a large number of parameters, however those are also subject to model uncertainties and impose correlations across the population parameter space~\cite{Golomb:2022bon,Edelman:2022ydv,Callister:2023tgi}. 
Detailed model checking remains an essential ingredient of population constraints.

For end-to-end event simulation and population recovery such as Fig.~\ref{fig:betaPlusDoubleGaussianTrace}, we \textit{a priori} know what the ``true" underlying astrophysical distribution is. 
However, when dealing with real GW observations, this is, of course, not the case. 
We therefore diagnose the bias seen in Fig.~\ref{fig:betaPlusDoubleGaussianTrace} using only information available to us when dealing with real observations.
To do so, we use posterior predictive checks (PPCs) that examine the predictive accuracy of the inferred models via its ability to predict future data that are consistent with current observations. 
PPCs are ubiquitous in the field of GW population analyses~\cite{Fishbach:2019ckx,Fishbach:2020qag,Callister:2022qwb,O3a-pop,O3b-pop}.

We now look at the results from the \textsc{Beta+DoubleGaussian} model presented in Sec.~\ref{sec:measuring}: a case in which a population model is theoretically a good fit, but cannot find the underlying distribution accurately.
A PPC for the \textsc{LowSpinAligned} population\footnote{We highlight the \textsc{LowSpinAligned} population throughout Sec.~\ref{sec:ppcs} as it displays the largest bias under the \textsc{Beta+DoubleGaussian} model.} is plotted in Fig.~\ref{fig:PPCs}. 
Specifically, we plot the spin parameters predicted by the fitted model against those of the observed events.
The ``predicted" (horizontal axis) and ``observed" (vertical axis) draws and are generated as follows: 
\begin{enumerate}
    \item Draw one sample from the posterior for the \textsc{Beta+DoubleGaussian} hyper-parameters.
    \item Draw one sample from the \textit{detectable}~\cite{Callister:2023tgi,Essick:2023upv} $\chi_i$ and $\cos\theta_i$ distribution corresponding to this hyper-parameter. This is the \textit{predicted} draw.
    \item Draw one sample from one individual-event posterior in the catalog, reweighted to the population from Step 1, as described in Appendix~\ref{appendix:reweighting}. This is the \textit{observed} draw.
    \item Repeat 70 times for the O3-like catalog, or 300 times for the O4-like catalog.
\end{enumerate}
The predicted and observed values are sorted and plotted against each other, generating one trace in Fig.~\ref{fig:PPCs}. 
We repeat this procedure 100 times to generate a collection of traces.
If we have perfectly measured the true underlying distribution and in the limit of infinitely many observations, the traces should be an exact diagonal. 
For a number of finite observations, the \textit{average} of the traces should be diagonal~\cite{Sinharay:2003,Bayarri:2007,Fishbach:2019ckx,Callister_reweighting,Miller:2020zox}.
As the number of observed events increases, the spread of the traces around the diagonal should decrease. 

For all spin parameters shown, the traces on average \textit{do} follow the diagonal, \textit{even though the measured population does not match the truth}. 
The 300 event case (navy) traces are more tightly clustered around the diagonal than the 70 event case (pink), as expected.
That the traces average to the diagonal but we know the fit is poor indicates that this class of PPC, although widely used in GW population analyses, is not a sufficient diagnostic of model mismatch or inaccurate population inference in this case.

We continue to investigate the conditions under which PPCs succeed or fail by next turning to the second case discussed previously: that in which a population model is an \textit{intrinsically} bad fit to the underlying astrophysical distribution.
We again perform population inference of simulated data, now deliberately using a model that \textit{cannot} reproduce the injected spin distributions. 
In particular, we will adopt a model in which cosine tilts are described only as a single Gaussian, which we call the \textsc{Beta+Gaussian} model and is given analytically in Appendix~\ref{appendix:betaPlusGaussian}. 
This population model \textit{is} capable of reproducing the $\chieff$ distribution, but at the level of component spins cannot capture the bimodality present in the \textsc{MediumSpin} and \textsc{LowSpinAligned} populations.

We here wish to isolate the effect of a bad model, \textit{without} having to worry about the shortcomings of inference per Fig.~\ref{fig:betaPlusDoubleGaussianTrace}.
As such, this time we do not perform full signal injection and parameter estimation, but instead produce mock spin magnitude and cosine tilt posteriors, allowing us to better control and understand the interplay between individual-event and population-level measurement uncertainty.
We assume these mock-posteriors to be Gaussian distributed with width $\sigmeas$. 
Our procedure for generating these mock-posteriors is detailed in Appendix~\ref{appendix:mock_posteriors}.

Results from conducting hierarchical inference using the Gaussian mock-posteriors are shown in Fig.~\ref{fig:mockPopsPPCs}. 
The left-hand column (blue), shows results from individual-event spin posteriors with $\sigmeas=0.1$, between $\sim\,$1-5 times more informative than the \textsc{Bilby}-produced $\cos\theta$ posteriors, which averages to $\sigmeas=\trueSigmaMeas$.
In the right-hand column (orange) are plotted results from more-realistic individual-event measurement error of $\sigmeas=0.5$; the center column (pink) is an intermediate case of $\sigmeas=0.3$.

The inferred $\cos\theta$ distributions for the \textsc{LowSpin- Aligned} population under the \textsc{Beta+Gaussian} model are plotted in the top row of Fig.~\ref{fig:mockPopsPPCs}. 
For each of the three different individual-event measurement errors, the traces are clustered tightly, meaning that the inferred population is precisely measured, even though the model is not a good fit to the underlying population. 
The \textsc{Beta+Gaussian} model is, in essence, doing its job: even with large individual event uncertainty, it identifies the mean and the over-all width of the distribution very well, even though it cannot capture the full underlying bimodal structure.

We again ask: if we did \textit{not} know the injected distribution, would we have been able to tell that this model is insufficient?
Going further, in the case that we \textit{can} tell a PPC fails, we are looking for an estimate of how we should ammend our population model to better fit the truth.
Beyond just inspecting the diagonality of PPCs by eye, we can calculate the fraction of events over/under-predicted by our model across parameter space using the slopes of the PPC traces. 
If the slope of a PPC trace is steeper (shallower) than the diagonal, then the model is predicting more (fewer) events in that region of parameter space than are observed. 
To find the slopes of each trace as a function of each parameter of interest, we perform linear regression in a small region around each point on a grid spanning that parameter.
The fractions of each spin parameters under-predicted for each simulated population is then the fraction of traces with slopes shallower than the diagonal (i.e.~$<1$).
If the model is a good fit to the data, the fraction under-predicted should be consistent with 0.5. 

PPCs and the corresponding fraction of events under-predicted are shown in the middle and bottom rows of Fig.~\ref{fig:mockPopsPPCs} respectively. 
Errors on the fraction under-predicted are calculated by repeating the PPC procedure ten times, and calculating the mean (crosses) and variance (shaded region) of the results.
For the $\sigmeas=0.1$ case, the PPC is inconsistent with the diagonal, meaning that here we \textit{can} identify that inferred distribution under the \textsc{Beta+Gaussian} model is not a good fit.  
The fraction of events under-predicted is correspondingly inconsistent with 0.5. 
For $\cos\theta \lesssim 0.25$ and $\cos\theta \gtrsim 0.75$, the fraction under-predicted is greater than 0.5, meaning that the model ubiquitously under-predicts the population in this region of parameter space. 
Between $0.25 \lesssim \cos\theta \lesssim 0.75$, the fraction is less than 0.5, meaning here the model over-predicts.
Looking at the top left corner of Fig.~\ref{fig:mockPopsPPCs}, we can see that this is exactly the case. 
These results hint at how we could improve the $\cos\theta$ model: to find the truth we should allow the model to predict more events at alignment and anti-alignment, i.e., include a bimodality. 

As the individual-event measurement uncertainty increases (left to right), the PPCs become more consistent with the diagonal, and correspondingly the fractions become consistent with 0.5. 
By realistic measurement uncertainty, we lose our ability to diagnose inconsistency between the underlying and measured populations using PPCs. 
A crucial step in generating PPCs is the reweighting of individual-event posteriors to the inferred population. 
If individual-event posteriors are sufficiently uninformative, then this process yields reweighted posteriors that are all essentially identical to the measured population. 
Thus, the ``observed" and ``predicted" draws will be the same, and the PPCs will be on average diagonal.
\emph{This type of PPC is therefore insufficient for weakly informative parameters as the reweighted posterior is dominated by the population rather than the individual-event likelihood.}
We propose that alternative model-checking procedures must, therefore, be developed and utilized for diagnosing model bias and misspecification for poorly-measured BBH parameters such as spin components.

\section{Comparison to past work on Hierarchical Inference with Spin-Precession}
\label{sec:discussion}

To our knowledge, our study includes the first full, end-to-end individual-event and population-level GW injection campaign for multiple distinct populations of spin magnitudes \textit{and} tilt angles of BBHs. 
Past studies performing injection-recovery campaigns for spin populations either use different waveform models and sampling implementations, and/or consider cases with reduced complexity compared to ours. 
Our work is in consistent with past findings, as described below.

\citet{Talbot:2017yur} investigated the measurability of the spin tilt angle distributions alone using an astrophysically-motivated model assuming that some fraction of BBHs form in isolated binaries, while the rest form dynamically.
They performed an injection and recovery campaign for spin tilts where they measured the fraction of binary mergers with preferentially-aligned versus isotropically distributed tilts, and the typical degree of spin misalignment for each BH. 
In their study, all simulated binaries share the same masses, distance, and spin magnitudes, chosen to be similar to LIGO's first event GW150914~\cite{GW150914}.
Using the waveform model \textsc{IMRPhenomPv2}~\cite{Hannam:2013oca} and nested sampling implemented in \textsc{LALInference}~\cite{LALinference}, they found that they \textit{are} able to constrain the parameters of the tilt-angle distributions for five different populations.
Although they \textit{do} sample over all fifteen BBH parameters during individual-event inference, \citet{Talbot:2017yur} is most similar to the follow-up studies we present in Appendix~\ref{appendix:reducing_complexity}: we too are able to better recover the underlying distributions for all three of our populations when the complexity of the explored parameter space is reduced (see, e.g.~Fig.~\ref{fig:fixed_chi_or_cost_dist}), on either an individual-event or population level.

In the context of searching for unresolved binary signals, \citet{Smith:2020lkj} also simulated and recovered BBH spin magnitude and tilt distributions. 
They looked at a single population, consistent with the LIGO/Virgo O1 and O2 observations~\cite{O2-pop}, and used the \textsc{IMRPhenomPv2} waveform model~\cite{Hannam:2013oca} implemented in \textsc{Bilby} for individual-event inference. 
A crucial difference between this study and ours is the use of a selection function: as \citet{Smith:2020lkj} are looking at resolved \textit{and} unresolved binaries, they ignore selection effects entirely. 
Under these conditions, \citet{Smith:2020lkj} find that the spin magnitude and tilt angle distributions are both accurately measurable. 
Given that their simulated populations are most similar to our \textsc{MediumSpin} population, this is in agreement with our findings, as we are able to well constrain the \textsc{MediumSpin} population for both 70 and 300 events.
It is only when more complex distributions are introduced that our inference fails. 

Another point of comparison between our work and others' is on the subject of biased measurements from population model misspecification. 
In particular, other authors have also identified shortcomings of traditionally and widely used model checking techniques such as probability-probability plots (Appendix~\ref{appendix:PPplots}) and  posterior predictive checks (Sec.~\ref{sec:ppcs}). 
\citet{Biscoveanu:2021eht} discussed population model bias in the mass distribution of binary neutron star populations arising from misspecification of spin priors. 
As part of their work, they show that a P-P check on individual-event posteriors can pass but still lead to highly biased population inference. 
This is in agreement with our findings.

\section{Conclusions}
\label{sec:conclusions}

In this work, we investigated the measurability of the spin magnitude and tilt angle distributions of BBH populations via GW observations.
To see if realistic GW populations contain information about spin components or just the effective spin, we simulated three BBH populations that have the same underlying effective-spin distributions, but deliberately distinct spin magnitude and tilt distributions, and on them conducted individual-event and population-level parameter estimation.
We then turned to the question of whether mismatch between the injected and recovered spin magnitude and tilt distributions can be identified using only the individual-event and population-level data available to us, without knowledge of the true underlying population.
Our work focuses on the three questions posed in Sec.~\ref{sec:intro}, the answers to which we summarize below. 

\textit{(1) }\textbf{\textit{There is information in gravitational-wave signals beyond the effective-spin.}} 
As discussed in Sec.~\ref{sec:measuring}, we can tell that our three different populations have different spin magnitude and tilt distributions despite their having identical $\chieff$ distributions. 

\textit{(2) }\textbf{\textit{Measuring component spin distributions accurately is practically challenging.}}
Under standard, reviewed parameter estimation settings, we were able to accurately measure the spin magnitude and/or tilt angle distributions for some of the populations, but not all three.
The bimodal tilt distribution of the \textsc{LowSpinAligned} population proved especially resistant to being accurately constrained, even when using a population model that was inherently bimodal.
We employed a suite of verification methods to ensure the robustness of these results, which are enumerated in Sec.~\ref{sec:pe_bias}.
Notably, we find that the effective spin distribution \textit{is}, however, accurately measured no matter the degree of mismatch in the component spin model constraints; this is not true for the effective precessing spin $\chip$ which remains susceptible to biased spin magnitude and tilt inference.

Although we cannot say for certain, we hypothesize that the root of the mismatch between the recovered and underlying distributions is related to a lack of convergence in individual-event posteriors, the specifics of which are subtle enough to not present themselves via a standard P-P check (Appendix~\ref{appendix:PPplots}).
We do not claim that accurately recovering component spin population distributions is \textit{impossible} at current sensitivity, just challenging. 
Running {\tt Bilby} with more aggressive settings, while computationally costly, may very well fix the problems presented in this work. 
However, unlike our injection set, real observations do not come with an answer key.
If manually tuning sampler settings is a requirement to recover truth, we must be aware what these same errors could manifest in real LIGO/Virgo events.

\textit{(3)} \textbf{\textit{At current sensitivity, we cannot tell when measurements of component spin distributions are biased via the currently widely-used method of posterior predictive checks.}} 
Due to the fact that individual-event posteriors are extremely weakly informative about spin components, reweighting these posteriors to the inferred population distribution--a crucial step in conducting posterior predictive checks--yields individual-event measurements that are all nearly identical to the inferred population itself.
Nearly \textit{any} population model can seem like a good fit to poorly constrained data, as discussed in Sec.~\ref{sec:ppcs}. 

\citet{Fishbach:2019ckx} detailed different categories of posterior predictive checking for GW data. 
The most commonly used level is what we do in this work: performing consistency checks on the \textit{true} underlying parameters of the observed data versus predicted by the model. 
However, one can also conduct PPCs on the \textit{observed} parameters (e.g.~max likelihood parameters) of the data versus those predicted by the model. 
While checks on the true parameters are susceptible to the issues related to reweighting that we discuss in Sec.~\ref{fig:PPCs}, checks on observed parameters might be more constraining. 
However, they are far more computationally expensive to perform, as one must generate maximum likelihood values predicted by the model: this involves either running an optimization routine or conducting mock-parameter estimation on thousands of events. 
While trustworthy mock-parameter estimation exists for some parameters (e.g. masses)~\cite{Fishbach:2019ckx,Fairhurst:2023idl,Farah:2023vsc}, the imprint of spin magnitudes and tilt angles on data is more subtle and remains unincorporated into these algorithms. 
Developing different, more-informative methods of posterior predictive checking for poorly-constrained parameters such as spin is an essential topic of future work.

\section*{Data and Code Availability}

\noindent
The code used to produce all results presented in this paper can be found at \url{https://github.com/simonajmiller/measuring-bbh-component-spin}.
Our individual-event and hierarchical-inference posteriors samples can be shared upon request.


\acknowledgements

We thank Sylvia Biscoveanu, Jacob Golomb, Ethan Payne, and Colm Talbot for their extensive assistance with \textsc{Bilby} parameter estimation and other helpful comments on this work; Sophie Hourihane for essential insights about reweighting and probability-probability plots; and Salvatore Vitale for useful feedback and discussion
Furthermore, we extend thanks to Maya Fishbach, Reed Essick, Matthew Mould, Zoheyr Doctor, and Will Farr for their suggestions related to the range of verification methods performed in this paper.
Finally, we thank our anonymous referees for their feedback.

This work is supported by National Science Foundation grant No. PHY-2150027 as part of the LIGO Caltech REU Program, which funded ZK.
S.M and KC were supported by NSF Grant PHY-2110111 and NSF Grant PHY-2308770. 
TC is supported by the Eric and Wendy Schmidt AI in Science Postdoctoral Fellowship, a Schmidt Futures program.
The authors are grateful for computational resources provided by the LIGO Laboratory and supported by National Science Foundation Grants PHY-0757058 and PHY-0823459.

Software: \texttt{emcee}~\cite{emcee}, \texttt{bilby} (version 2.2.2)~\cite{bilby,Romero-Shaw:2020owr}, \texttt{dynesty} (version 2.1.2) ~\cite{dynesty}, \texttt{numpy}~\cite{numpy}, \texttt{scipy}~\cite{scipy}, \texttt{matplotlib}~\cite{matplotlib}, \texttt{seaborn}~\cite{seaborn}, \texttt{astropy}~\cite{astropy1,astropy2}, \texttt{jax}~\cite{jax}, \texttt{numpyro}~\cite{numpyro1,numpyro2}


\appendix

\section{Simulated Populations}
\label{appendix:pops}

We simulate three populations with the same $\chieff$ but different spin magnitude $\chi$ and tilt angle $\theta$ distributions. 
To generate the populations, we first choose distributions of the mass ratio $q$ and spin $z$-component $s_{i,z}\equiv\chi_i \cos\theta_i$ to be shared in common across all three populations; this ensures the same $\chieff$ distribution.
The mass ratio distribution corresponds to the median posterior value inferred with the \textsc{PowerLaw+Peak} model in Ref.~\cite{O3b-pop}, while for $s_{i,z}$ we select a Gaussian with mean $\szMean$ and standard deviation $\szStd$.
This yields a Gaussian-like $\chieff$ distribution with mean $\chieffMean$ and standard deviation $\chieffStd$.

To decompose $\chieff$ into component spins, we choose a different spin magnitude $\chi$ distribution for each population and then numerically calculate the resultant $\cos\theta$ distribution implied by $p(\chi)$ and $p(s_{i,z})$.
This procedure results in the three populations shown in Fig.~\ref{fig:populations}.
The $\chi$ distribution for the \textsc{HighSpinPrecessing} population is uniform between $s_{i,z}$ and $1$, i.e.,~$s_{i,z}$ values are drawn from the Gaussian described above and then a $\chi_i$ value is conditionally drawn based on each $s_{i,z}$.
For the \textsc{MediumSpin} (\textsc{LowSpinAligned}) population, each $\chi$ value is drawn from a Gaussian distribution about $s_{i,z}$, truncated on $0\leq\chi\leq1$, with a standard deviation of $\popTwoChiStd$ ($\popThreeChiStd$).
For each population, we assume $\chi_1$ and $\chi_2$ are identically but independently distributed, as are $\cos\theta_1$ and $\cos\theta_2$. 
Finally, each spin vector's azimuthal angle $\phi_i$ is drawn uniformly between $0$ and $2\pi$.
Due to the different $\chi$ and $\cos\theta$ distributions, the $\chi_p$ distributions of each population differ as well.\footnote{Different component spin distributions given identical $\chieff$ \emph{and} $\chi_\mathrm{p}$ distribution can only be achieved by relaxing the assumption of identically distributed component spin magnitudes and angles.
}

The astrophysical distribution of the remaining binary parameters is the same for all populations.
We inject primary masses drawn from the \textsc{PowerLaw+Peak} model~\cite{Talbot:2018cva} with all parameters, except for $m_\mathrm{min}$, fixed to their one-dimensional median values as found in Ref.~\cite{O3b-pop}: $\alpha=3.51$, $m_\mathrm{max} = 88.21$, $\lambda_\mathrm{peak} = 0.033$, $\mu_m = 33.61$, $\sigma_m = 4.72$, and $\delta_m = 4.88$ in the notation used therein.
The injected mass ratio distributions for all populations are described by a power law with slope $\beta_q=0.96$ (see Eq.~\ref{eqn:massratio_PL}), again the median inferred value from~\cite{O3b-pop}.
In the parameterization of the \textsc{PowerLaw+Peak} model, we use a population minimum mass of $m_\mathrm{min} = 6\,M_\odot$ instead of $5\,M_\odot$ to set the shape of the distribution.
We additionally impose a mass \textit{cut} of $8\,M_\odot$, as restricting to higher-mass events ensures shorter analysis times.
This mass cut effectively becomes the minimum mass, but we renormalize the distribution to keep the same \textit{shape} above the cutoff mass as it would with $m_\mathrm{min} = 6$.
Explicitly setting $m_\mathrm{min} = 8$ in the \textsc{PowerLaw+Peak} model would change the over-all shape of the distribution to be inconsistent with the desired results in Ref.~\cite{O3b-pop}.

Finally, the BBH merger density rate in the source frame evolves with respect to redshift $z$ as
    \begin{equation}
    R(z) \propto \frac{dV_c}{dz} \left(1+z\right)^{2.7}\,,
    \end{equation}
where $V_c$ is the comoving volume.
Distances are calculated from redshifts assuming the cosmology reported by the Planck 2013 survey~\cite{Planck:2013pxb}.
All other parameters are drawn uniformly from their respective physical range.
Mass and redshift distributions are plotted in Fig.~\ref{fig:mass_redshift_dists}.

\section{Individual-event Parameter Estimation}
\label{appendix:pe}

\begin{figure}
    \centering
    \includegraphics[width=0.47\textwidth]{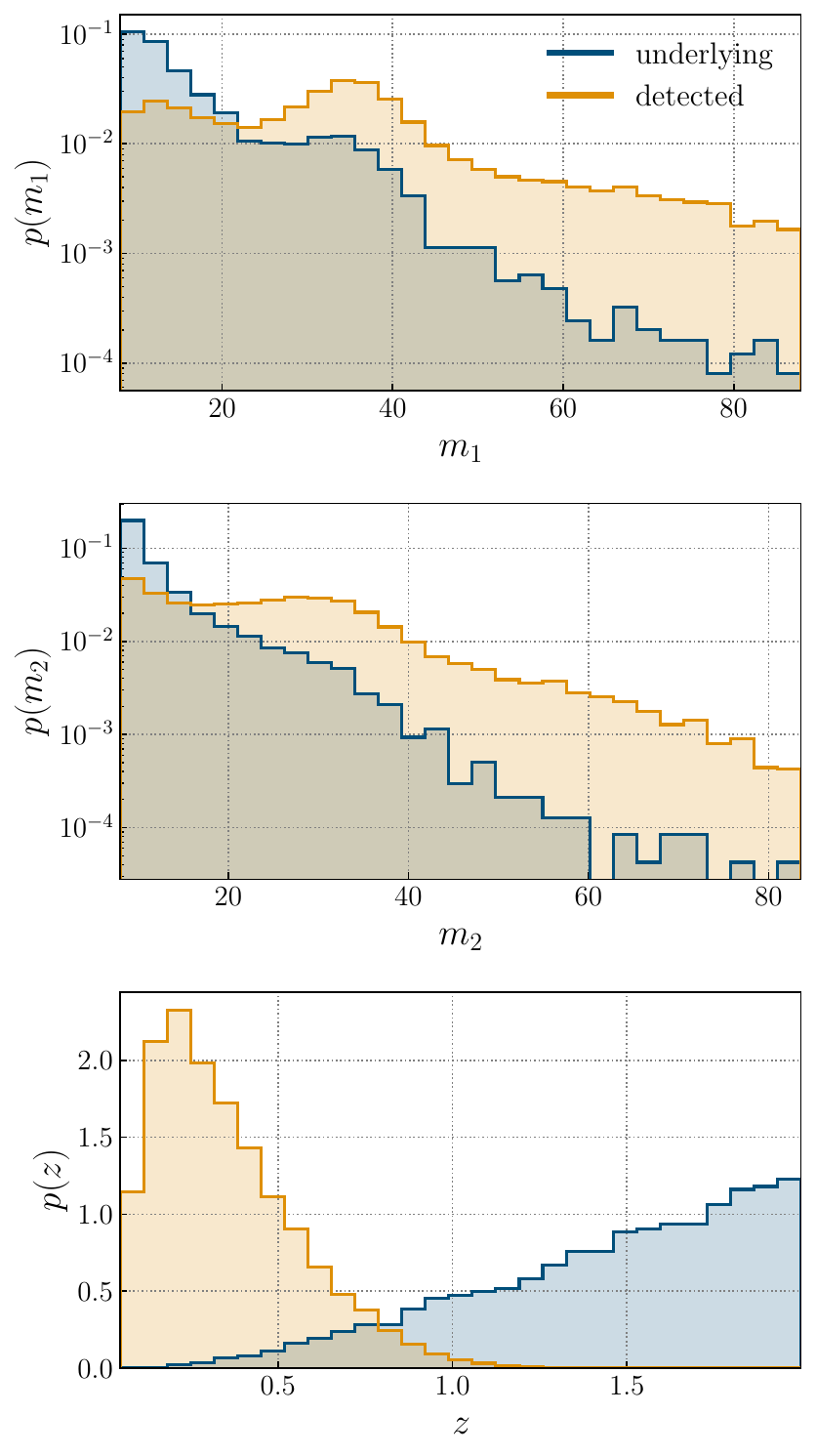}
    \caption{The underlying (navy) and detected (orange) distributions for source-frame primary mass $m_1$ (top), source-frame secondary mass $m_2$ (middle), and redshift $z$ (bottom) shared between all three simulated populations.}
    \label{fig:mass_redshift_dists}
\end{figure}

From each of the three astrophysical distributions described in Appendix~\ref{appendix:pops}, we draw $10^5$ events. 
We apply a network SNR~\cite{Thorne:1987,Finn:1992xs,Allen:2005fk,Essick:2023toz} cut of 10 in the LIGO Livingston, LIGO Hanford, and Virgo detector network using the ``O3 actual" power spectral densities (PSDs) provided in Ref.~\cite{O3_PSDs}, and select 300 detectable events.
Histograms of events from the underlying (navy) versus detectable (orange) mass and redshift distributions are shown in Fig.~\ref{fig:mass_redshift_dists}.

For each event, we simulate GW data with the \textsc{IMRPhenomXPHM} waveform model~\cite{Pratten:2020ceb} including a Gaussian noise realization and draw samples from the 15-dimensional posterior distribution for the binary parameters using the same waveform.
Specifically, we sample in detector frame component masses $m_1, m_2$, spin magnitudes $\chi_1,\chi_2$, spin tilt angles $\theta_1,\theta_2$, the azimuthal inter-spin angle $\phi_{12}$, the azimuthal cone precession angle $\phi_{JL}$, the luminosity distance $d_L$, the inclination angle between the total angular momentum and the line of sight of the observer $\theta_{JN}$, the right ascension $\alpha$, declination $\delta$, polarization angle $\psi$, and the time $t$ and orbital phase $\varphi$ at coalescence. 
We employ standard priors for all binary parameters~\cite{Romero-Shaw:2020owr}, although we use a targeted chirp mass~\cite{Peters:1964,Blanchet:1995} prior of $\pm 15 M_\odot$ about the injected value to reduce computational cost, which we verify does not affect results.

Simulated data assume a detector network of LIGO-Hanford, LIGO-Livingston~\cite{aLIGO}, and Virgo~\cite{aVirgo}, each at their O3 sensitivity~\cite{O3_PSDs} with a sampling rate of $2048~\mathrm{Hz}$.
We analyze data in the $15-921.6(=0.9 \times 2048/2 )~\mathrm{Hz}$ frequency range, assuming perfect knowledge of the detector calibration. 

We use the nested sampler \textsc{dynesty}~\cite{dynesty} as implemented in \textsc{Bilby}~\cite{bilby,Romero-Shaw:2020owr} under reviewed settings to stochastically sample from the individual-event posteriors. 
For sampler settings, we use \texttt{nlive = 1000}, \texttt{naccept= 60}, and \texttt{sample = "acceptance-walk"}.
Time marginalizaton is turned on, while distance and phase marginalization remain off. 
Post-facto, we apply an optimal SNR cut of 10 on the posterior samples for consistency with our selection criteria~\cite{Essick:2023upv}.

\section{Hierarchical Inference}
\label{appendix:inference}

The hierarchical inference framework used in our analysis to obtain posteriors distributions on the population parameters is implemented using the Python Markov-Chain Monte-Carlo package \texttt{emcee}~\cite{emcee}.
The likelihood $\mathcal{L}(\{d\}|\Lambda)$ that a catalog of $N_\mathrm{obs}$ GW events with data $\{d_i\}_{i=1}^{N_\mathrm{obs}}$ arises from an underlying population $\pi_\mathrm{pop}$ described by  parameters $\Lambda$ is given by~\cite{Loredo:2004nn,Mandel:2018mve,Thrane:2019,Vitale:2020aaz}
    \begin{equation}
    \mathcal{L}(\{d\}|\Lambda) \propto \prod^{N_\mathrm{obs}}_i
    \frac{
        \int d\lambda \,\mathcal{L}(d_i|\lambda) \pi_\mathrm{pop}(\lambda|\Lambda)}{\xi(\Lambda)}\,,
    \label{eq:likelihood}
    \end{equation}
where $\lambda_i$ are the parameters of the $i^\mathrm{th}$ event in the catalog (i.e.~spins, masses, etc).
In practice, we have access to the individual-event posteriors $p(\lambda_i|d_i)$ obtained with a default parameter-estimation prior $\pi_\mathrm{pe}(\lambda_i)$, rather than the event likelihood $\mathcal{L}(d_i|\lambda_i)$.
We thus write Eq.~\eqref{eq:likelihood} as
    \begin{equation}
    \mathcal{L}(\{d\}|\Lambda) \propto \xi(\Lambda)^{-N_\mathrm{obs}} \prod_i
        \int d\lambda \,\frac{p(\lambda|d_i)}{\pi_\mathrm{pe}(\lambda_i)}  \pi_\mathrm{pop}(\lambda_i|\Lambda)\,.
    \label{eq:likelihood-post}
    \end{equation}
Additionally, rather than $p(\lambda_i|d_i)$ itself, we have a discrete set of $N_i$ independent samples $\{\lambda_{i,j}\}_{j=1}^{N_i}$ drawn from $p(\lambda_i|d_i)$.
Using the standard procedure, we approximate the integral of Eq.~\eqref{eq:likelihood-post} via a Monte Carlo average,
    \begin{equation}
    \mathcal{L}(\{d\}|\Lambda) \propto \xi(\Lambda)^{-N_\mathrm{obs}}
        \prod_i
        \frac{1}{N_i}\sum_{j=1}^{N_i} \frac{\pi_\mathrm{pop}(\lambda_{i,j}|\Lambda)}{\pi_\mathrm{pe}(\lambda_{i,j})}\,.
    \label{eq:likelihood-mc}
    \end{equation}

The detection efficiency 
    \begin{equation}
    \xi(\Lambda) = \int d\lambda \,\pi_\mathrm{pop}(\lambda|\Lambda) P_\mathrm{det}(\lambda)\,,
    \label{eq:detection-efficiency}
    \end{equation}
is the fraction of events that we would successfully detect if the population with parameters $\Lambda$ is the true underlying population. Here, $P_\mathrm{det}(\lambda)$ is the probability that an individual event with parameters $\lambda$ is detected. 
As with the population likelihood, we calculate the detection efficiency with a Monte Carlo average.
Given $N_\mathrm{inj}$ injected signals drawn from some reference distribution $p_\mathrm{inj}(\lambda)$, the detection efficiency is
    \begin{equation}
    \xi(\Lambda) = \frac{1}{N_\mathrm{inj}} \sum_{i=1}^{N_\mathrm{fnd}} \frac{\pi_\mathrm{pop}(\lambda_i|\Lambda)}{p_\mathrm{inj}(\lambda_i)}\,,
    \label{eq:detection-efficiency-mc}
    \end{equation}
where the sum is over the $N_\mathrm{fnd}$ injections that pass the detection criteria.
We generate the set of ``found" injections over which the Monte Carlo average is calculated in the same way that we produced catalogs of events in Appendix~\ref{appendix:pe}. The reference $p_\mathrm{inj}(\lambda)$ follows the true mass and redshift distribution (Appendix~\ref{appendix:pops}; Fig.~\ref{fig:mass_redshift_dists}), but is uniform in spin magnitudes and isotropic in spin tilts such that we can resolve features across the full underlying spin distribution.  
As in Appendix~\ref{appendix:pe}, our detection criterion is an optimal SNR greater than 10 using the waveform \textsc{IMRPhenomXPHM}~\cite{Pratten:2020ceb} in the LIGO Livingston, LIGO Hanford, and Virgo network at O3 sensitivity~\cite{O3_PSDs}.
We acknowledge that the optimal SNR is not a strictly accurate estimate of selection effects on real data as it is solely a function of source parameters, not detector noise. 
However, this approach remains formally self-consistent as long as we apply the same optimal SNR cut on posterior samples, as explained in~\citet{Essick:2023upv}.

Following~\citet{Farr:2019}, we account for uncertainty in the Monte Carlo integral by demanding that the effective number of independent samples  
    \begin{equation}
        \Neff(\Lambda) \equiv \frac{\left[ \sum_{i=1}^{N_\mathrm{fnd}} w_i(\Lambda)\right]^2}{\sum_{i=1}^{N_\mathrm{fnd}} \left[ w_i (\Lambda)\right]^2} \geq 4\,N_\mathrm{obs}\,,
        \label{eq:Neff}
    \end{equation}
where the weights $w_{i}$ between the population distribution and parameter estimation prior are defined as
\begin{equation}
    w_i(\Lambda) = \frac{\pi_\mathrm{pop}(\lambda_i|\Lambda)}{p_\mathrm{inj}(\lambda_{i})}\,,
    \label{eqn:Neff_weights}
\end{equation}
evaluated on the parameters of the \textit{found} injections. 
This procedure rejects samples from regions of parameter space in which there are not sufficient injections to accurately probe. 
We use 200,000 injections to calculate $\xi$.
Our results never rail against the $\Neff$ cut of Eq.~\eqref{eq:Neff}, so we do not believe it affects our results. 
For further investigation, we perform a set of analyses \textit{without} including spins when calculating $\Neff$ (see Appendix~\ref{appendix:misc}), under which our conclusions do not change.

In addition to including a cut on effective samples from the selection function, one can also impose a cut on the \textit{per-event} effective samples of the posteriors used in calculating the hierarchical likelihood. 
Here, instead of evaluating Eq.~\ref{eqn:Neff_weights} on the found injections, it is evaluated on the \textit{posterior samples} for every event. 
If any events in the catalog have an effective sample number below some threshold, the corresponding $\Lambda$ sample is tossed. 
In this work, we do not include any per-event $\Neff$ cuts in the sampling of $\mathcal{L}(\{d\}|\Lambda)$, but calculate them post-facto as a check of Monte Carlo convergence (see e.g.~Fig.~\ref{fig:diff_catalog_instantiations}).
Other tests of Monte Carlo convergence are discussed in \cite{Talbot:2023pex}.

\section{Spin population models}
\label{appendix:models}

We recover the simulated populations with two models.
Generically, we factorize the population models as
\begin{equation}
\begin{split}
    \pi_\mathrm{pop}(\lambda | \Lambda) & 
    = p(m_1 | \Lambda )\,p(m_2 | m_1, \Lambda )\,p(z | \Lambda) \times \\
       & \, p( \chi_1 | \Lambda)\, p(\chi_2 |  \Lambda) \, p(\cos\theta_1| \Lambda)\, p( \cos\theta_2| \Lambda)\,,
\end{split}
\end{equation}
meaning, aside from the masses $m_1$ and $m_2$, there are no correlations in the population. 
Our two parameterized models for the spin magnitude $\chi$ and tilt angle $\theta$ are described below. More details about the two models are provided in Table~\ref{tab:component-models}.
Spin magnitudes $\chi_i$ and tilt angles $\theta_i$ are assumed identically and independently distributed.

During hierarchical inference, we fix the distributions of primary mass $m_1$ and redshift $z$ to truth, as described in Appendix~\ref{appendix:pops}.
To account for possible correlations between spins and mass ratio, although no underlying correlation was injected, we follow~\citet{Callister:2021fpo} and simultaneously infer the distribution of binary mass ratios and spins using a secondary mass distribution of
   \begin{equation}\label{eqn:massratio_PL}
    p(m_2|m_1) \propto m_2^{\beta_q} \quad \left(m_\mathrm{min} \leq m_2 \leq m_1\right)\,,
    \end{equation}
where the power-law index $\beta_q$ is a free parameter with a Gaussian prior of $\mathcal{N}(0,3)$.
The true underlying distribution has $\beta_q = 0.96$.
For all other individual-event parameters, we take population distributions identical to the priors used during the original \textsc{Bilby} parameter estimation.
Most notably for this analysis, azimuthal spins are distributed uniformly $\phi_i \in [0, 2\pi)$.
All parameters aside from masses, redshift, and spin magnitudes and tilt angles are thus excluded from hierarchical inference. 

\begin{table*}[]
    \centering
    \renewcommand{\arraystretch}{1.25}%
    \begin{tabular}{l c|c|c|c}
         Model Name & $p(\chi)$ \qquad \qquad \quad $p(\cos \theta)$ & Parameter & Prior & Comments\\
         \hline
         \hline
         &&& \\[-12pt]
         \multirow{4}{*}{\textsc{Beta+Gaussian}}  &    
         \multirow{4}{*}{\includegraphics[width=40mm]{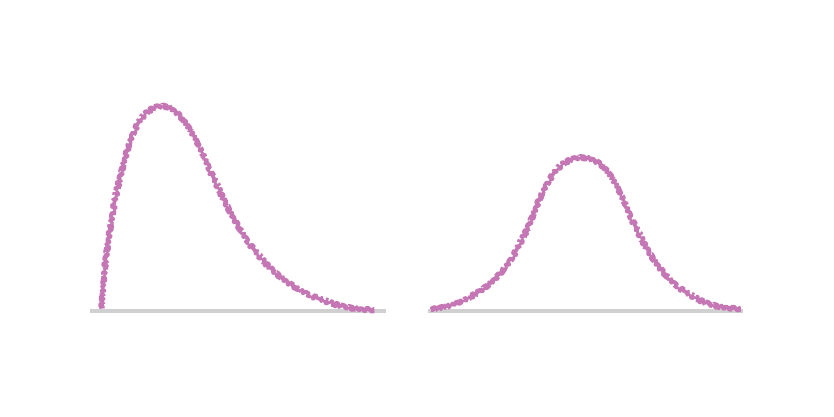}} & $\mu_\chi$ & U(0,1) & \multirow{4}{4cm}{Component spin model that cannot reproduce the simulated populations; used to study model misspecification} \\
         & & $\sigma_\chi$ & U(0.07,0.5) & \\
         & & $\mu_{\theta}$ & U(-1,1) & \\
         & & $\sigma_{\theta}$ &  U(0.16,0.8) & \\[2pt]
         \hline
         \multirow{5}{*}{\textsc{Beta+DoubleGaussian}}&    
         \multirow{5}{*}{\includegraphics[width=40mm]{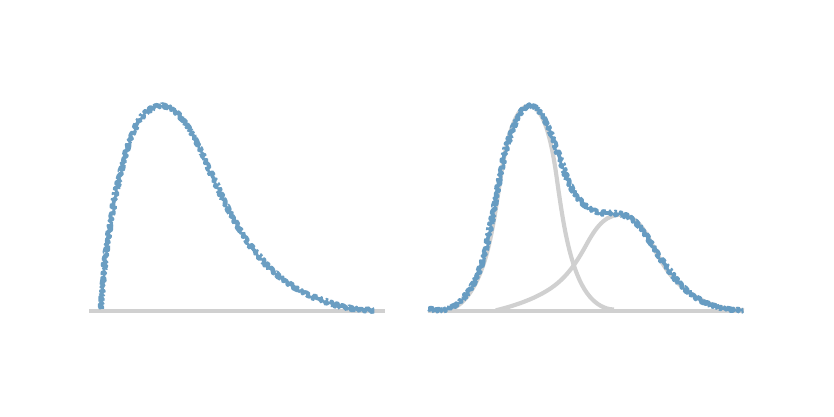}} &  $\mu_\chi$ & U(0,1)  & \multirow{5}{4cm}{Component spin model that can reproduce the simulated populations; used to study the amount of information available in component spins. See Appendix~\ref{appendix:misc} about methods of breaking the degeneracy between the two Gaussian components.} \\
         & & $\sigma_\chi$ & U(0.07,0.5) & \\
         & & $\mu_{\theta,1}$ & U(-1,1) & \\
         & & $\sigma_{\theta,1}$ &  U(0.16,0.8) & \\
         & & $\mu_{\theta,2}$ & U(-1,1) & \\
         & & $\sigma_{\theta,2}$ &  U(0.16,8) & \\
         & & $f$ &  U(0,5) & \\[2pt]
         \hline
         \hline
    \end{tabular}
    \caption{Details about the two component spin models we employ. Columns give the model names, an example $\chi$ and $\cos \theta$ plot, the parameters they depend on, the parameter priors, and some brief comments. The notation U$(a,b)$ means a prior uniform between $a$ and $b$. For $p(\chi)$ in both models, we impose an additional prior requirement that the beta distribution shape parameters $\alpha, \beta>1$, see Eq.~\eqref{eq:mu-sigma-to-alpha-beta}, making the distribution non-singular at the boundaries. Full expressions for $p(\chi)$ and $p(\cos\theta)$ can be found in Eqs.~\eqref{eq:beta},~\eqref{eq:gaussian}, and~\eqref{eq:doubleGaussian}.}
    \label{tab:component-models}
\end{table*}
%

\subsection{\textsc{Beta+Gaussian}}
\label{appendix:betaPlusGaussian}

Following the \textsc{Default} model in Ref.~\cite{O3b-pop}, we assume that spin magnitudes $\chi_i$ are identically and independently distributed according to a Beta distribution
\begin{equation}
    p(\chi_i|\alpha,\beta) = \frac{\chi_i^{\alpha-1} (1-\chi_i)^{\beta-1}}{B(\alpha,\beta)}\,,
\label{eq:beta}
\end{equation}
where $B(\alpha, \beta)$ is the Beta function which ensures that the distribution is normalized to unity on $0\leq\chi\leq1$.
Instead of sampling in the shape parameters $\alpha$ and $\beta$, we sample in the more familiar mean $\mu_\chi$ and standard deviation $\sigma_\chi$ which are related to $\alpha$ and $\beta$ by 
\begin{equation}
    \alpha =\mu_\chi \nu ~ , \quad
      \beta = (1-\mu_\chi) \nu \,,
\label{eq:mu-sigma-to-alpha-beta}
\end{equation}
where 
\begin{equation}
    \nu = \frac{\mu_\chi(1-\mu_\chi)}{\sigma_\chi^2} - 1 \,\,.
\end{equation}
We adopt uniform priors on $\mu_\chi$ and $\sigma_\chi$ and impose an additional cut such that $\alpha, \beta \geq 1$ to keep the distribution bounded. 
This cut enforces $p(\chi_i|\alpha,\beta) = 0$ at $\chi_i=0$ and $1$.
Aside from the \textsc{HighSpinPrecessing} population at $\chi=1$, this assumption is valid, and even there the spin model captures the over-all distribution's shape.

For the tilt-angle distribution, we adopt a truncated, normalized Gaussian distribution
\begin{equation}
    p(\cos\theta_i | \mu_{\theta}, \sigma_{\theta} ) =  \mathcal{N}_{[-1,1]}(\cos\theta_i | \mu_{\theta}\,, \sigma_{\theta})
\label{eq:gaussian}
\end{equation}
on the interval $-1\leq\cos\theta\leq1$, and fit for the mean $\mu_\theta$ and standard deviation $\sigma_\theta$.

\subsection{\textsc{Beta+DoubleGaussian}}
\label{appendix:betaPlusDoubleGaussian}

The \textsc{Beta+DoubleGaussian} model uses the same spin magnitude distribution as the \textsc{Beta+Gaussian}, as given in Eq.~\eqref{eq:beta} and explained thereafter.
The tilt angle distribution is here instead given by a mixture of two truncated normalized Gaussians
\begin{multline}
    p(\cos\theta_i | \mu_{\theta,1}, \sigma_{\theta,1}, \mu_{\theta,2}, \sigma_{\theta,2}, f) = \\ f \mathcal{N}_{[-1,1]}(\cos\theta_i | \mu_{\theta,1}, \sigma_{\theta,1}) \\ + (1-f) \mathcal{N}_{[-1,1]}(\cos\theta_i | \mu_{\theta,2}, \sigma_{\theta,2})\,,
\label{eq:doubleGaussian}
\end{multline}
to capture the multimodality of the some of the underlying distributions.
We measure the means $\mu_{\theta,1}$, $\mu_{\theta,2}$  and standard deviations $\sigma_{\theta,1}$, $\sigma_{\theta,2}$ of the two Gaussians, and the mixing fraction $f$ between them. 
We impose $\mu_{\theta,1} \leq \mu_{\theta,2}$ to distinguish between the two components.

\section{Detailed Hierarchical Inference Results}
\label{appendix:cornerplots}

\begin{figure*}[p]
\centering
    \includegraphics[width=\linewidth]{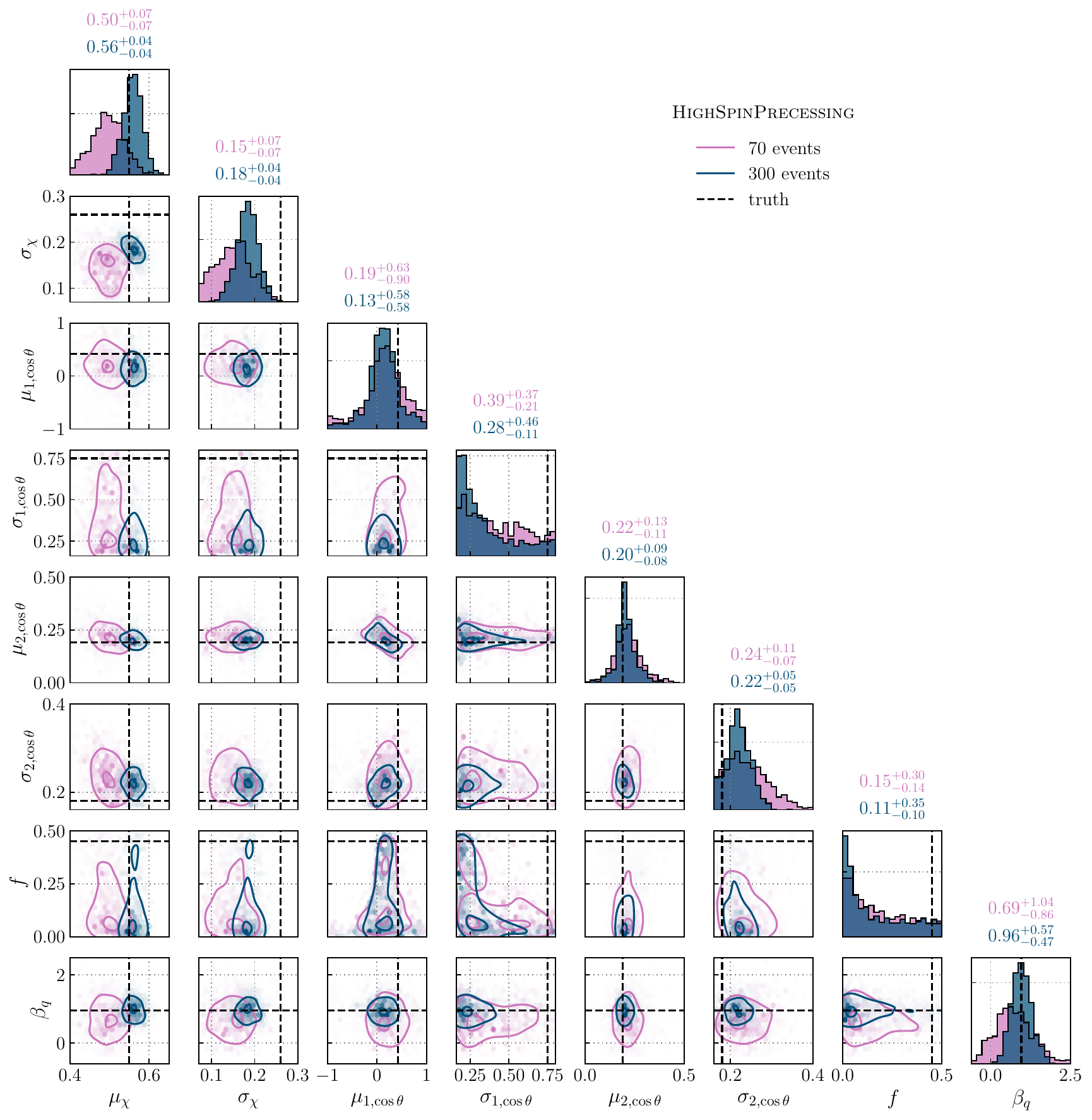}
    \caption{
    The posterior distributions on the hyper-parameters of the spin magnitude and tilt angle distributions under the \textsc{Beta+DoubleGaussian} model for the \textsc{HighSpinPrecessing} population for 70 (pink) and 300 (navy) event catalogs.
    The labels above each one-dimensional posterior give the medians and 90\% credible intervals on each hyper-parameter for the two different catalog sizes, while the contours in each two-dimensional posterior denote the 50\% and 90\% credible regions.
    See Table \ref{tab:component-models} for descriptions the hyper-parameters and their priors. 
    Black dashed lines labeled ``truth" represent the theoretical best-fit parameters for the population under the \textsc{Beta+DoubleGaussian} model, as calculated using a least-squared fit on 50,000 draws from the population.
    }
    \label{fig:cornerplots1}
\end{figure*}

\begin{figure*}
\centering
    \includegraphics[width=\linewidth]{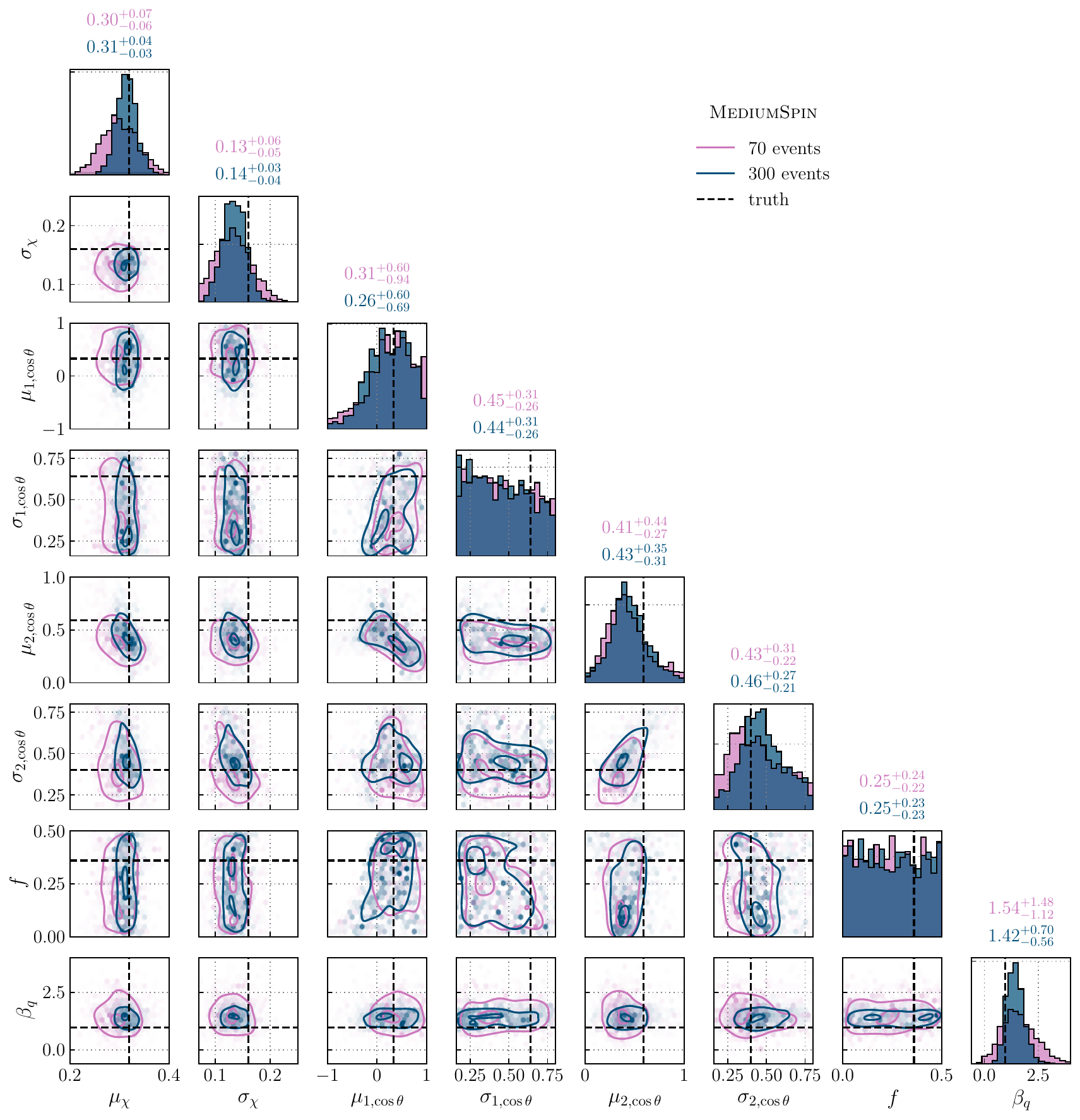}
    \caption{
    Same as Fig.~\ref{fig:cornerplots1} but for the \textsc{MediumSpin} population.
    This population is recovered by the \textsc{Beta+DoubleGaussian} model without bias.
    }
    \label{fig:cornerplots2}
\end{figure*}

\begin{figure*}
\centering
    \includegraphics[width=\linewidth]{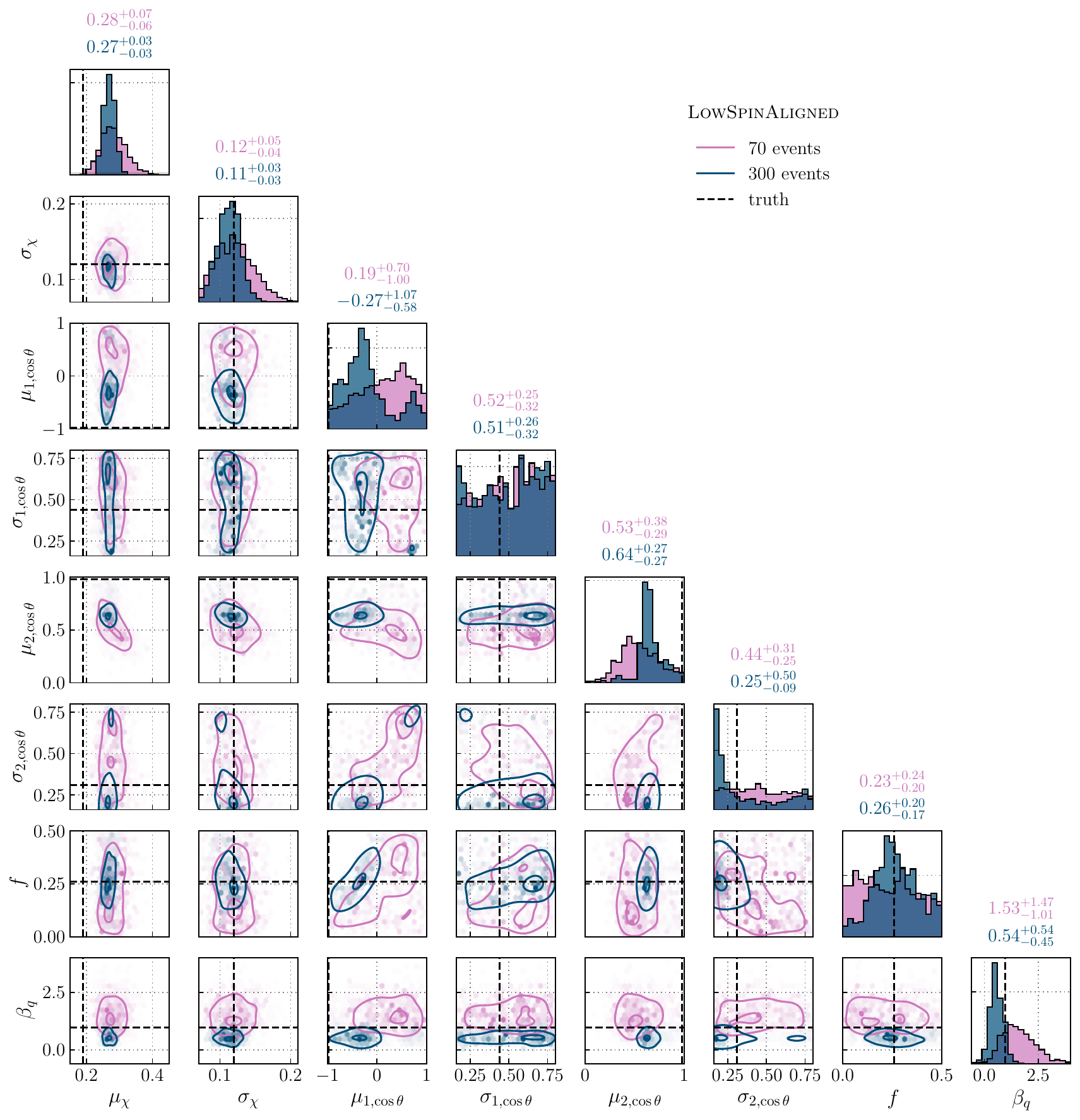}
    \caption{
    Same as Figs.~\ref{fig:cornerplots1} and \ref{fig:cornerplots2} but for the \textsc{LowSpinAligned} population.
    This population is recovered by the \textsc{Beta+DoubleGaussian} model with considerable bias.
    }
    \label{fig:cornerplots3}
\end{figure*}

We here show full posteriors on the hyper-parameters for the \textsc{Beta+DoubleGaussian} component spin population model, as given in Eqs.~\eqref{eq:beta} and \eqref{eq:doubleGaussian}, for the \textsc{HighSpinPrecessing} (Fig.~\ref{fig:cornerplots1}), \textsc{MediumSpin} (Fig.~\ref{fig:cornerplots2}), and \textsc{LowSpinAligned} (Fig.~\ref{fig:cornerplots3}) populations. 
Results for 70 (pink) and 300 (navy) event catalogs are shown in each figure. 
These posteriors are compared against non-linear least-squares fit parameters (black dashed) calculated from 50,000 draws per population, representing the best possible fit for the true underlying distributions within the \textsc{Beta+DoubleGaussian} model.
Population distributions generated from draws from these posteriors are plotted in Fig.~\ref{fig:betaPlusDoubleGaussianTrace}.
For all three populations, as expected, including more events makes hyper-parameter measurements more \textit{precise}.
However, adding more events does not necessarily make the results more \textit{accurate}. 

The hyper-parameters of the \textsc{HighSpinPrecessing} population (Fig.~\ref{fig:cornerplots1}) are recovered with minimal bias. 
While the mean $\mu_\chi$ of the spin magnitude distribution of the \textsc{HighSpinPrecessing} population is very well constrained, its width $\sigma_\chi$ is slightly underestimated in the case of both the 70 and 300 event catalogs. 
The means $\mu_{i,\cos\theta}$ are also accurately constrained.
The widths of the tilt angle distributions also seem to be under-estimated, but per Fig.~\ref{fig:betaPlusDoubleGaussianTrace}, the actual shape of the distribution converges on the truth. 
This is because -- due to the allowed bimodality -- different combinations of hyper-parameters can lead to the same unimodal distribution.

The \textsc{MediumSpin} population (Fig.~\ref{fig:cornerplots2}), on the other hand, is reconstructed very accurately by the \textsc{Beta+DoubleGaussian} model for both the 70 and 300 event catalogs.
Each ``true" hyper-parameter either falls within the 90\% measured credible region.
This is also reflected in Fig.~\ref{fig:betaPlusDoubleGaussianTrace} -- the black traces representing truth are enclosed by the 90\% credible envelopes for both $\chi$ and $\cos\theta$.

Finally, the width of the spin magnitude distribution for the \textsc{LowSpinAligned} population (Fig.~\ref{fig:cornerplots3}) is accurately constrained, but its mean $\mu_\chi$ is over-estimated. 
The most striking failure of the \textsc{Beta+DoubleGaussian} model is its inability to identify the bimodality of the \textsc{LowSpinAligned} population's tilt angle distribution, for either 70 or 300 events.
Aside from the mixing fraction $f$, none of the tilt-angle distribution hyper-parameters' posteriors are consistent with truth at the 90\% level. 

For all three populations, the power law slope for the mass distribution is recovered within 90\% credibility about the injected value of $\beta_q = 0.96$. 
Masses are recovered without bias by our hierarchical inference procedure; we only encountering biases when fitting for the spin populations. 

\section{Reweighting individual event posteriors}
\label{appendix:reweighting}

Given a set of discrete sample from an individual-event posterior $p(\lambda|d_i)$ calculated with prior $\pi_\mathrm{pe}(\lambda)$ and discrete samples of hyper-parameters describing a population distribution $\pi_\mathrm{pop}(\lambda | \Lambda)$, we can reweight the individual-event posterior to the inferred population using a two-step algorithm~\cite{Miller:2020zox, Callister_reweighting}. 
First, randomly select a hyper-parameter sample from the population distribution $\Lambda_i \in \{\Lambda\}$ and calculate the following weights for each individual-event posterior sample $\lambda_j$: 
\begin{equation}
    w_j \propto \frac{\pi_\mathrm{pop}(\lambda_j | \Lambda_i)}{\pi_\mathrm{pe}(\lambda_j)} \,\,.
\end{equation}
Second, select one sample $\lambda_j \in \{\lambda_j\}$ subject to the weights $w_j$.
Repeat this process to build up a set of samples from a reweighted individual-event posterior. 
This procedures ensures that events are not double-counted during weighting~\cite{Callister_reweighting}.

\section{Verification of methods}
\label{appendix:verification}

To explore the origin of the bias observed in Fig.~\ref{fig:betaPlusDoubleGaussianTrace}, we perform a number of explorations that we elaborate upon in subsequent subsections.
In Appendix~\ref{appendix:PPplots}, we generate probability-probability (P-P) plots for reweighted individual event \textsc{Bilby} posteriors, which return unbiased. 
In Appendix~\ref{appendix:mock_posteriors}, we turn to hierarchical inference and explore a range of simulated Gaussian individual-event spins posteriors, rather than ones generated via stochastic sampling with \textsc{Bilby}.
We also reduce the complexity of the parameter spaces explored on both an individual-event and population level. 
In Appendix~\ref{appendix:reducing_complexity}, we present results where we fix various combinations of parameters to their true values in both the individual-event and hierarchical inference levels. 
We here also discuss using a less complex waveform -- \textsc{IMRPhenomXP} rather than \textsc{IMRPhenomXPHM} -- which excludes higher order modes, in individual-event sampling. 
Appendix~\ref{appendix:alt-codes} uses an alternate hierarchical inference code, implemented in \textsc{Numpyro} instead of \textsc{emcee}, and 
Appendix~\ref{appendix:masses} shows results from simultaneously inferring for the mass and redshift distributions along with the spins.
In Appendix~\ref{appendix:rates}, we look at hierarchically-inferred rates across parameter space rather than probability density functions to ensure that the normalization is not obscuring the results. 
Finally, other miscellaneous checks for the hierarchical inference framework and implementation are: excluding selection effects in spin; trying different methods of breaking the degeneracy in the double-Gaussian tilt distribution; and looking at different 70-event catalog instantiations. 
Plots showing these results can be found in Appendix~\ref{appendix:misc}.

\subsection{Probability-probability (P-P) plots}
\label{appendix:PPplots}

A crucial assumption of hierarchical inference is that the input individual-event posteriors are themselves reliable. 
To test this assumption and ensure that the stochastically-sampled \textsc{Bilby} individual-event posteriors (see Appendix~\ref{appendix:pe}) are indeed unbiased, we perform the common diagnostic check of generating a probability-probability (P-P) plot~\cite{Cook:2006,Talts:2018}. 

A P-P plot is generated by performing parameter estimation on events with parameters distributed according to their individual-event priors, in Gaussian noise. 
The percentiles, or credible intervals (CI), at which the injections fall in their resultant one-dimensional, marginalized posterior distribution shall be uniformly distributed if parameter estimation is unbiased. 
In our case, where the injected distribution does \textit{not} match the priors used in parameter estimation, reweighting (see Appendix~\ref{appendix:reweighting}) to the injected distribution must be performed as a post-processing step.  
Specifically, we apply an optimal SNR cut of 10 to the posteriors, and then reweight to the underlying population; this procedure is analogous to \textit{not} applying any SNR cut and reweighting to the detected distribution. 

\begin{figure*}
\centering
    \includegraphics[width=\linewidth]{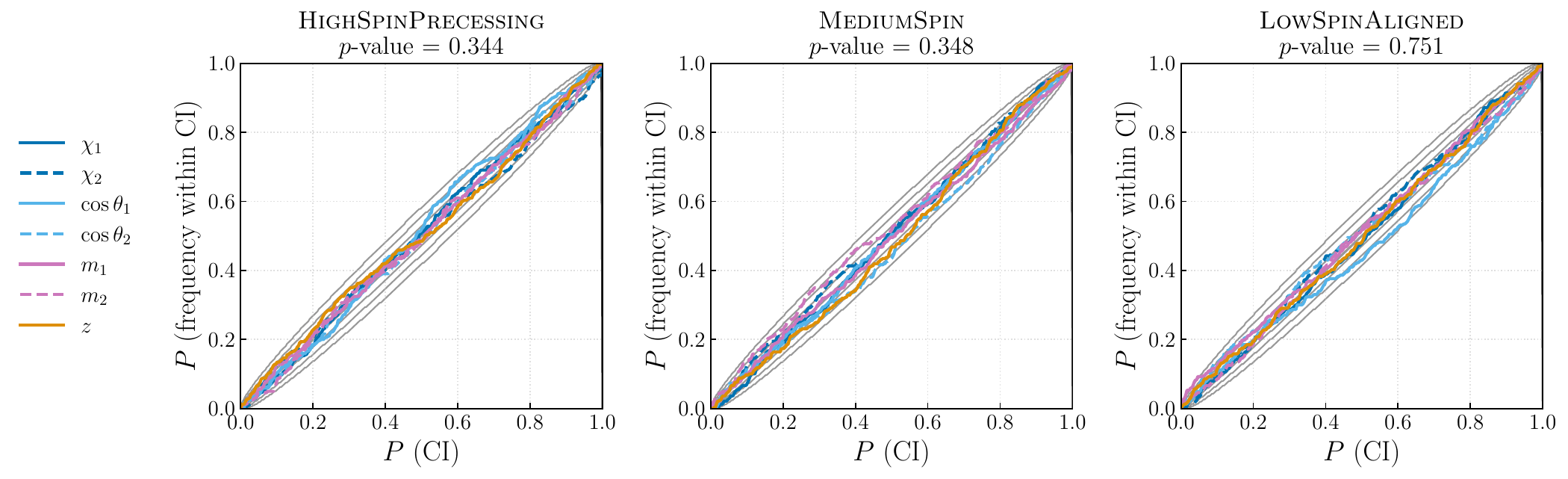}
    \caption{
    Probability-probability (P-P) plots for spin magnitudes (dark blue), spin tilt angles (light blue), masses (pink) and redshifts (orange) for each simulated population (from left to right: \textsc{HighSpinPrecessing}, \textsc{MediumSpin}, \textsc{LowSpinAligned}). 
    For 300 events per population, confidence intervals (CIs; horizontal axis) are plotted against the fraction of events for which the true, injected value is recovered in that CI, stochastically sampled using the \textsc{Bilby} implementation of the nested sampler \textsc{Dynesty}.
    The 1-, 2-, and 3-$\sigma$ regions for 300 events are plotted in gray; all parameters stay within the 3-$\sigma$ region, corresponding to the outermost gray lines. 
    The $p$-values for each of the three populations are greater than the threshold for 7 parameters ($\sim0.143$), indicating that the P-P test is passed.
    }
    \label{fig:PPplots}
\end{figure*}

P-P plots for spin magnitudes, spin tilt angles, masses, and redshifts for the 300 injections per simulated population are shown in Fig.~\ref{fig:PPplots}.
On the horizontal axis, the (sorted) CIs are plotted. 
The vertical axis shows the cumulative density function (CDF) of these CIs, i.e.~the frequency at which each CI occurs.
This should be a diagonal line with a slope of 1 in the case of infinitely many injections: e.g.~20\% of the time, the injection should fall within the lower 20\% CI of its posterior.
In the case of finitely many injections, these CDFs should roughly fall within a 3-$\sigma$ region around the diagonal, the width of which is a function of the number of events injected, as indicated by the gray lines in Fig.~\ref{fig:PPplots}.

To check if the \textsc{Bilby} posteriors pass the P-P test, we look at the $p$-values\footnote{We calculate $p$-values using a Kolmogorov-Smirnov test.} that each set of $y$-axis values shown in Fig.~\ref{fig:PPplots} is uniform.
Then, we take the $p$-values \textit{of} these $p$-values, which should also be uniformly distributed if the sampling error is random. 
The $p$-values (listed in the titles of Fig.~\ref{fig:PPplots}) for each of the three simulated populations is above the threshold of randomness expected from the 7 parameters plotted ($1/7\sim0.143$), indicating that the \textsc{Bilby} posteriors pass the P-P test.

Though a necessary check, diagonal P-P plots are not a sufficient condition for reliable individual-event posteriors. 
As also seen in~\cite{Biscoveanu:2021eht}, a sampling algorithm can pass a P-P test but still result in biased hierarchical inference recovery beyond the 3-$\sigma$ level. 
In Fig.~\ref{fig:cornerplots3}, for example, the truth for the means of the spin magnitude distribution and both modes of the spin tilt distribution all lie outside of the 90\% credible interval for the recovered values. 
It remains unclear in our case whether such discrepancies between the true and recovered populations are due to individual-event sampling issues that are not picked up by the P-P test, as was the case in~\cite{Biscoveanu:2021eht}, or further unknown biases.

\subsection{Simulated Gaussian individual-event spin posteriors}
\label{appendix:mock_posteriors}

To better understand the relation between individual-event and population-level measurement uncertainty, we generate a series of simulated individual-event spin magnitude and tilt angle posteriors for each of the same 300 GW events per population that we stochastically sample in \textsc{Bilby}. 
We take these mock posteriors to be Gaussian distributed with width $\sigmeas$. 

First, we generate a series of mock posteriors without any underlying spin-spin correlations with the following steps. 
For each of the 300 injections per population, 
\begin{enumerate}
    \item Take the true, injected value of each spin parameter
    $$\lambda_\mathrm{true}\in\{ \chi_1, \chi_2, \cos\theta_1, \cos\theta_2 \}\,,$$
    and from it draw an observed maximum likelihood value $\lambda_\mathrm{obs}$  from the Gaussian distribution $\mathcal{N}(\lambda_\mathrm{true}, \sigmeas)$ with mean $\lambda_\mathrm{true}$ and width $\sigmeas$.
    \item Draw $N$ samples from $\mathcal{N}_{[a,b]}(\lambda_\mathrm{obs}, \sigmeas)$ where $N$ is the number of samples in the \textsc{Bilby} posterior for the injection of interest and $\mathcal{N}_{[a,b]}$ is a Gaussian distribution \textit{truncated} on $[a,b]$. For spin magnitude this truncation is between $[0,1]$, and for the cosine tilt angle $[-1,1]$.
\end{enumerate}
Specifically, we look at cases where $\sigmeas = 0.1$, $0.3$, and $0.5$, as shown in Fig.~\ref{fig:PPCs}.
In all cases, we keep the \textsc{Bilby} mass and redshift posteriors. 
Moreover, the simulated and \textsc{Bilby} posteriors all have the same number of samples per event.

To simulate a more realistic case, we also generate a set of mock Gaussian posteriors that do include underlying inter-spin correlations, with the same covariance as \textsc{Bilby} individual-event posteriors. 
For each injection, we first find the covariance of the corresponding four-dimensional \textsc{Bilby} posterior for $\{ \chi_1, \chi_2, \cos\theta_1, \cos\theta_2 \}$.
We then generate a mock four-dimensional spin posterior with that same covariance using the procedure enumerated above: from truth, draw an observed maximum likelihood value; then generate a posterior by sampling a truncated Gaussian centered at that observed value. 
The only difference is, instead of separately generating each one-dimensional posterior for magnitudes and tilts, we generate a four-dimensional posterior that includes correlations.  

\begin{figure*}
\centering
    \includegraphics[width=\linewidth]{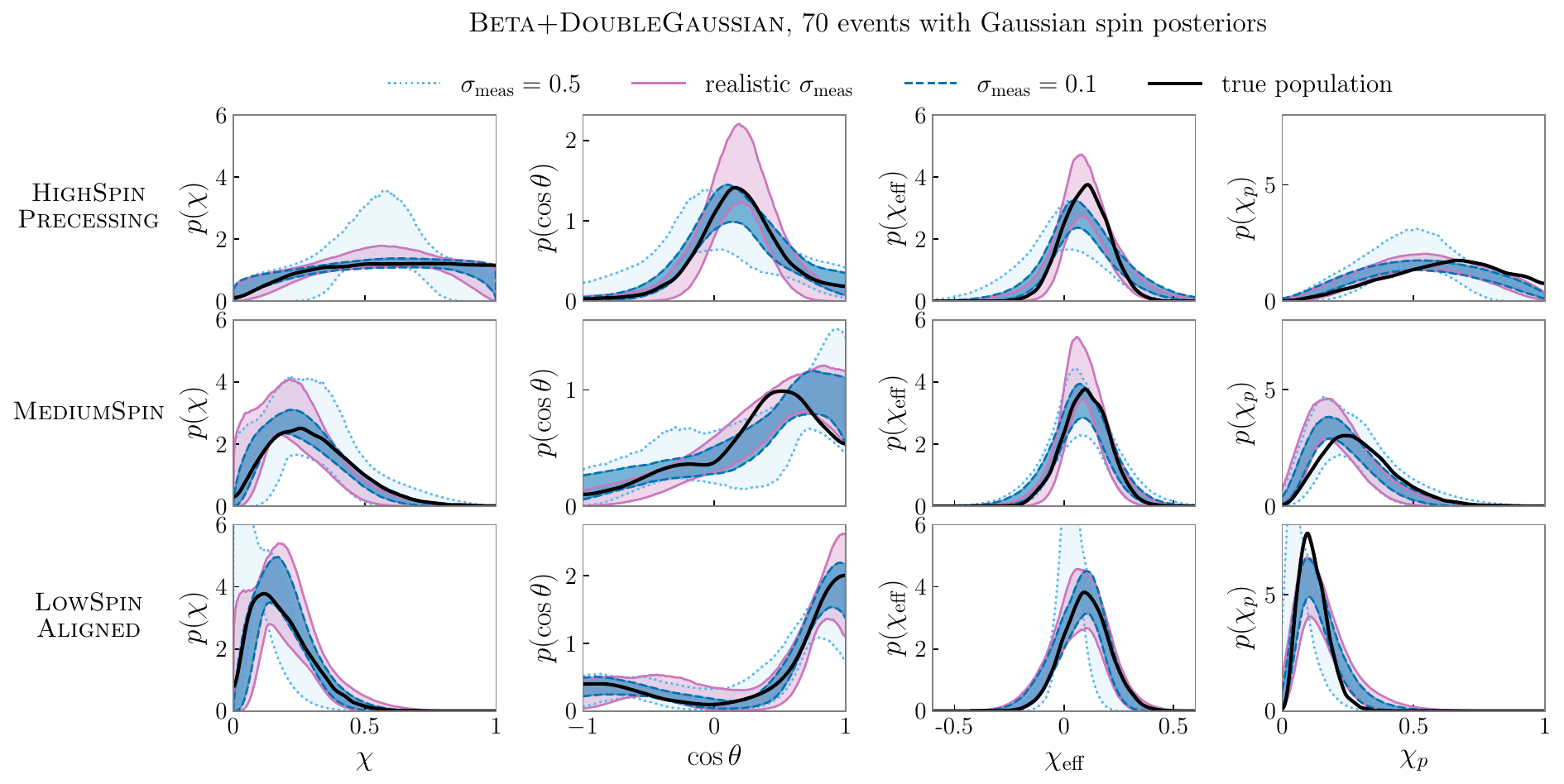}
    \caption{
     Inferred distributions obtained with the \textsc{Beta+DoubleGaussian} model for spin magnitude $\chi$, spin tilt $\cos\theta$, effective $\chieff$ spin, effective precessing spin $\chip$, for the three simulated populations with \textit{simulated} Gaussian individual-event spin posteriors.
     Results are shown from 70 event catalogs with per-event spin magnitude and tilt measurement error $\sigmeas = 0.5$ (light blue dotted), $\sigmeas = 0.1$ (blue dashed), and with realistic $\sigmeas$ and inter-parameter correlations taken from the \textsc{Bilby} posteriors (pink solid).
     The shaded region denotes 90\% of the probability, while the black solid line corresponds to the true population.
    }
    \label{fig:mockPopsBetaDoubleGaussian}
\end{figure*}

The \textsc{Beta+DoubleGaussian} population model \textit{is} able to recover the underlying populations when using the simulated Gaussian posteriors, as seen in Fig.~\ref{fig:mockPopsBetaDoubleGaussian}.
This is true for the most-informative individual-event mock-posteriors ($\sigmeas=0.1$; light blue dotted), the least informative ($\sigmeas=0.5$; blue dashed), and the most realistic (the posteriors with correlations, labeled ``realistic $\sigmeas$"; pink solid).
In all three cases, the true population lies within the 90\% credible interval of the recovered region. 
As the measurement error decreases, the constraints get tighter around truth.

\subsection{Reducing complexity of the explored parameter spaces}
\label{appendix:reducing_complexity}

In this section, we present two simplified scenarios for individual-event sampling in \textsc{Bilby}, then one simplified scenario for population inference in \textsc{emcee}. 

\begin{figure*}
\centering
    \includegraphics[width=0.9\linewidth]{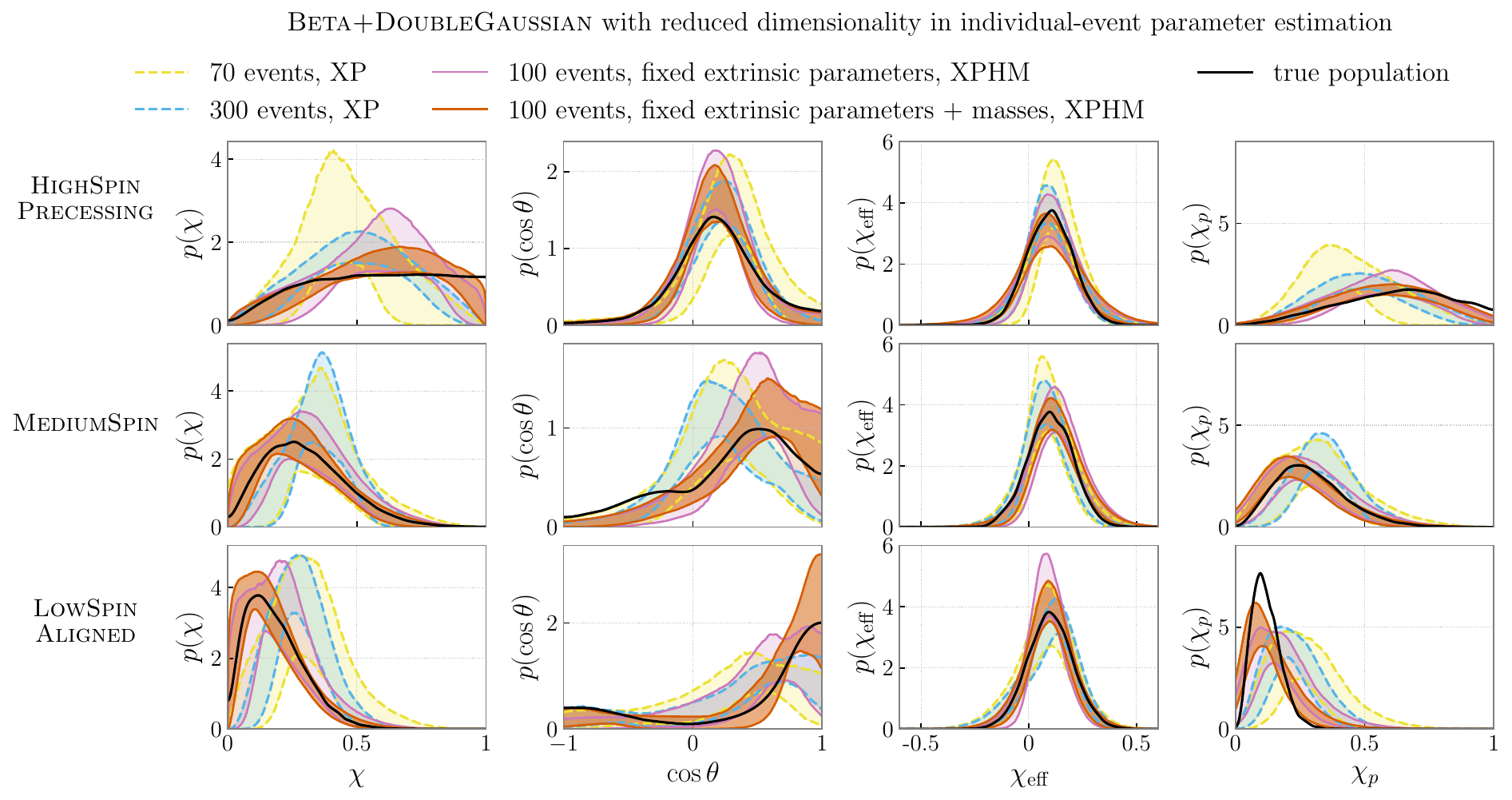}
    \caption{
     Inferred distributions obtained with the \textsc{Beta+DoubleGaussian} model for spin magnitude $\chi$, spin tilt $\cos\theta$, effective $\chieff$ spin, effective precessing spin $\chip$, for the three simulated populations with some form of reduced complexity when the sampling of individual-event posteriors.
     The dashed lines show population results with the \textsc{IMRPhenomXP} waveform (used for both injection and recovery), which excludes higher order modes, for 70 (yellow) and 300 (blue) event catalogs. 
     The solid lines shown populations results from 100-event catalogs using individual-event posteriors calculated with \textsc{IMRPhenomXPHM}, but with various parameters fixed to their true values rather than sampled over. 
     In pink (``fixed extrinsic''), we have fixed the extrinsic parameters (i.e.~everything aside from masses and spins) to their true values. 
     In orange (``fixed extrinsic + masses''), we further restrict by \textit{only} sampling over spin magnitudes and tilts. 
     Of these variations, only the ``fixed extrinsic + masses'' individual-event posteriors yield population constraints that are improved from those shown in Fig.~\ref{fig:betaPlusDoubleGaussianTrace}.
    }
    \label{fig:4D_and_8D_and_XP}
\end{figure*}

On the side of individual-event inference, we first reduce complexity by reducing the number of parameters sampled over. 
The results shown in the main text use posteriors where all fifteen dimensions of parameter-space are sampled over (see Appendix~\ref{appendix:pe}). 
To simplify, we first conduct parameter estimation on all the same injections, but fixing their extrinsic parameters (i.e.~everything aside from masses and spins) to the true, injected values. 
This yields the population constraints shown in the pink solid lines in Fig.~\ref{fig:4D_and_8D_and_XP} (for 100 events), labeled ``fixed extrinsic parameters." 
We then simplify further and additionally fix the masses and spin azimuthal angles to truth, generating the orange solid lines in Fig.~\ref{fig:4D_and_8D_and_XP} and labeled ``fixed extrinsic parameters + masses." 
Notably, the ``fixed extrinsic + masses'' individual-event posteriors yield population constraints that are significantly improved from those shown in Fig.~\ref{fig:betaPlusDoubleGaussianTrace}.
Since sampling convergence is more challenging as the dimensionality of the explored parameter-space increases, the trend we observe suggests that convergence might at least partially contribute to the bias.

Next, we return to sampling over all parameters (masses, spins, and extrinsic), but this time with a simpler waveform model: \textsc{IMRPhenomXP}~\cite{Pratten:2020ceb}. 
Coming from the same family as \textsc{IMRPhenomXPHM}, the \textsc{IMRPhenomXP} model does not contain higher order modes, which help break degeneracies between BBH parameters.
The yellow (light blue) dashed lines in Fig.~\ref{fig:4D_and_8D_and_XP} show the population constraints from the same 70 (300) events as Fig.~\ref{fig:betaPlusDoubleGaussianTrace} but with individual-event posteriors sampled with \textsc{IMRPhenomXP}. 
The recovered \textsc{HighSpinPrecessing} and \textsc{MediumSpin} populations have a worse mismatch with the truth than in Fig,~\ref{fig:betaPlusDoubleGaussianTrace}, but the \textsc{LowSpinAligned} population is recovered marginally better. 
Higher order modes become more important to accurately constrain BBH parameters as the degree of spin precession increases. 
Thus, the fact that the \textsc{HighSpinPrecessing} population is the worst constrained by \textsc{IMRPhenomXP} is consistent with our understanding of the utility of higher order modes.
We emphasize that these findings are unrelated to waveform systematics: we always inject and recover with the \textit{same} waveform model. 

\begin{figure}
\centering
    \includegraphics[width=\linewidth]{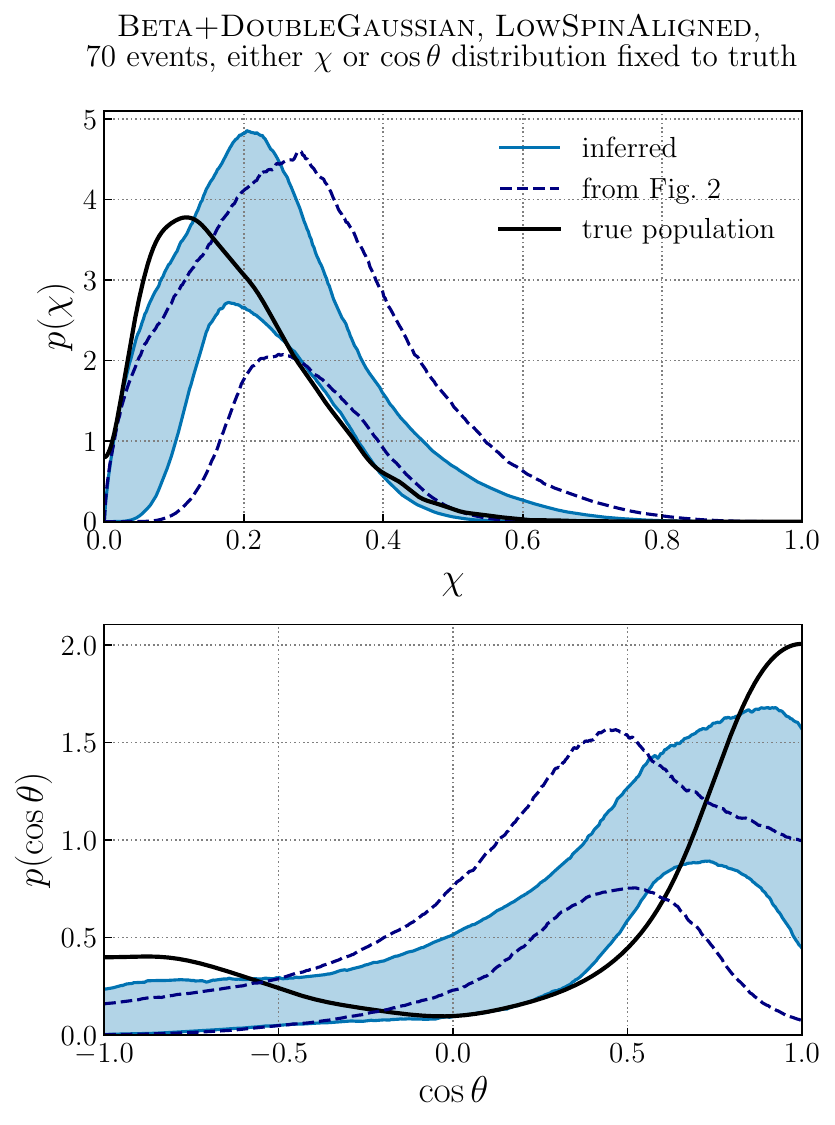}
    \caption{
     (\textit{Top}) Inferred spin magnitude $\chi$ distribution (blue shaded) for 70 events from the \textsc{LowSpinAligned} population under the \textsc{Beta+DoubleGaussian} population model with the tilt-angle distribution fixed to truth (black line in the bottom subplot). 
     (\textit{Bottom}) Inferred cosine of the tilt angle $\cos\theta$ distribution for the same 70 events and model with the spin magnitude distribution fixed to truth (black line in the top subplot).
     In both subplots, 90\% credible intervals for the distributions inferred by fitting for \textit{both} the spin magnitude and tilt distributions simultaneously, as shown in Fig.~\ref{fig:betaPlusDoubleGaussianTrace}, are shown in navy dashed lines.
    }
    \label{fig:fixed_chi_or_cost_dist}
\end{figure}

On the population level, to reduce the complexity of the sampling, we conduct analyses where we fit for \textit{only} the spin magnitude \textit{or} the tilt angle distribution, while fixing the other to it's true injected value. 
In theory, this could help identify if one or the other of these parameters was the driving factor for the mismatch between the true and recovered populations seen in Fig.~\ref{fig:betaPlusDoubleGaussianTrace}.
The inferred spin magnitude distribution for the \textsc{LowSpinAligned} population under the \textsc{Beta+DoubleGaussian} model with the tilt distribution fixed to truth is shown in blue in the top panel of Fig.~\ref{fig:fixed_chi_or_cost_dist}; the bottom panel shows the inverse. 
These recoveries are indeed better than those in Fig.~\ref{fig:betaPlusDoubleGaussianTrace} (plotted in navy dashed lines for comparison), i.e.~the mean of the $\chi$ distribution and mean of the larger sub-population of the $\cos\theta$ distribution are both more accurate. 
However, even in this much simplified version, the  \textsc{Beta+DoubleGaussian} model again fails to recover the truth, and in particular still shows no signs of bimodality in the tilt distribution.

\subsection{Hierarchical analysis with independent codes}
\label{appendix:alt-codes}

It is always possible that our poor recovery of component spin distributions is simply due to an unidentified error in the code used to perform hierarchical inference.
As a safeguard against this possibility, we have repeated the hierarchical analysis of the \textsc{HighSpinPrecessing}, \textsc{MediumSpin}, and \textsc{LowSpinAligned} populations using a second, distinct body of code.
This alternate analysis code was developed entirely independently, and furthermore relies on a different stochastic sampler:
whereas our main hierarchical inference results are obtained using \textsc{emcee}, this alternative performs inference using \textsc{numpyro}~\cite{numpyro1,numpyro2}, a probabilistic programming library implemented with \textsc{jax}~\cite{jax}.
The \textsc{numpyro}-based code produces results nearly identical to those obtained with our \textsc{emcee}-based code.
This implies that our results are not attributable to an unidentified error, unless that same error was independently introduced into two bodies of code created by two different analysts.

\subsection{Simultaneously fitting the mass and redshift distributions}
\label{appendix:masses}

Yet another source of potential bias we investigate is the choice to fix, rather than fit, the binary black hole primary mass and redshift distributions.
In principle, we do not expect significant covariance between the inferred primary mass, redshift, and component spin distributions; our simulated astrophysical populations have no underlying correlations between mass, redshift, and spin. 
At the same time, inferred component spins are expected to correlate strongly with the \textit{mass ratio} distribution, which in turn can depend systematically on the choice of primary mass distribution~\cite{Ng:2018neg,Biscoveanu:2021eht}.
Furthermore, it is known that assumptions regarding spin magnitudes can at times affect inference of the high-redshift rate of black hole mergers~\cite{O3a-pop,O3b-pop}.
Given these possibilities, it is possible that fixing the presumed mass and redshift distributions (even fixing them to the \textit{correct values}, as we have done) introduces bias into our spin measurements.

\begin{figure*}
\centering
    \includegraphics[width=\linewidth]{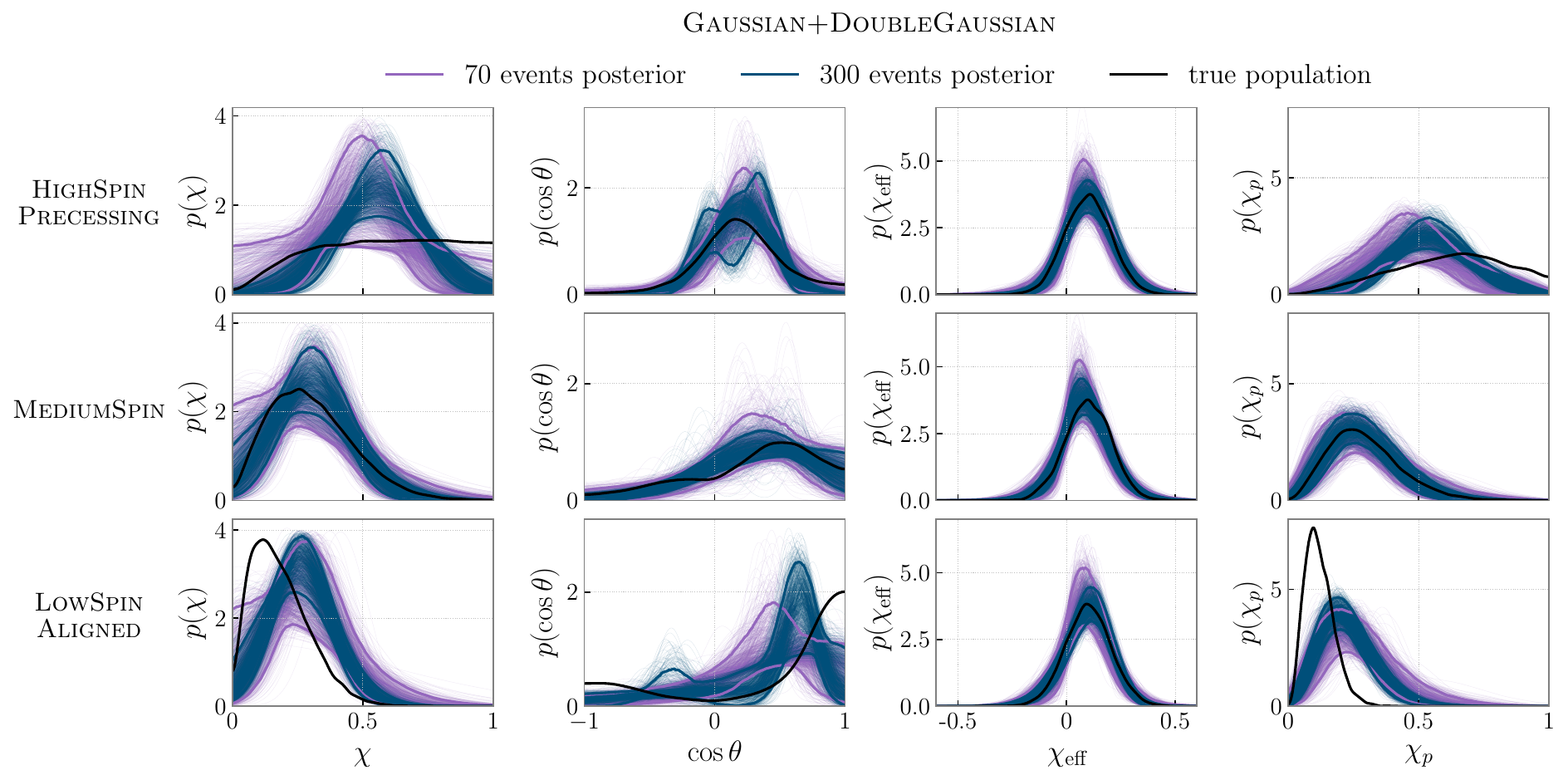}
    \caption{
    As in Fig.~\ref{fig:betaPlusDoubleGaussianTrace}, but now when additionally inferring the mass and redshift distributions of the three simulated populations in conjunction with their spin distributions.
    These results are furthermore produced with an entirely independent body of code, using \textsc{numpyro} rather than \textsc{emcee} to stochastically sample the population likelihood.
    Despite these differences, the results are nearly identical to those in Fig.~\ref{fig:betaPlusDoubleGaussianTrace}, indicating that the difficult recovery of injected component spin distributions is due neither to our choice to fix the mass and redshift distributions in the main text, nor to unidentified errors in our hierarchical inference code.
    }
    \label{fig:spins-with-masses}
\end{figure*}

To check this, we repeat our inference but now hierarchically inferring the black hole mass and redshift distributions alongside the component spin distributions.
We model primary masses following the \textsc{PowerLaw+Peak} model~\cite{O3a-pop,O3b-pop} and assume that the merger rate density evolves with redshift as $(1+z)^\kappa$ for some parameter $\kappa$.
This model also uses a slightly different spin magnitude model: a truncated Normal distribution instead of a Beta distribution.
We perform this inference using the alternative \textsc{numpyro}-based code introduced in Appendix~\ref{appendix:alt-codes} above.
The spin distributions inferred in this case are shown in Fig.~\ref{fig:spins-with-masses}.
The results are extremely similar to those in Fig.~\ref{fig:betaPlusDoubleGaussianTrace}.
As before, we recover the \textsc{HighSpinPrecessing} and \textsc{MediumSpin} component spin distributions reasonably well, but do not successfully measure the \textsc{LowSpinAligned} distributions.
In this latter case, we miss (or misplace) the bimodality inherent in $\cos\theta$ and, accordingly, systematically overestimate component spin magnitudes.
Once more, though, the $\chi_\mathrm{eff}$ distribution is well-recovered in all three cases.

\subsection{Recovering rates and spins simultaneously}
\label{appendix:rates}

\begin{figure*}
\centering
    \includegraphics[width=\linewidth]{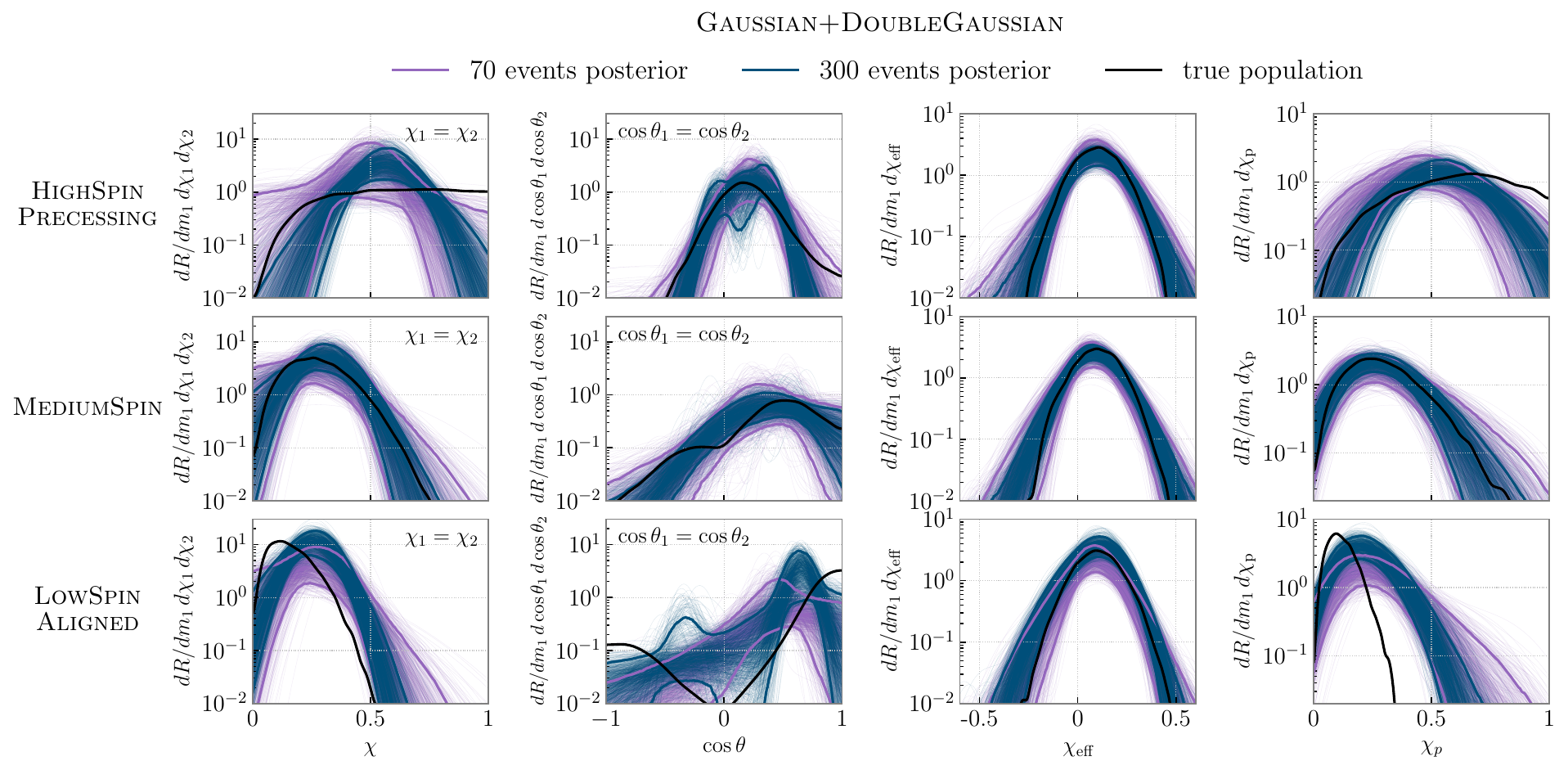}
    \caption{
    As in Fig.~\ref{fig:betaPlusDoubleGaussianTrace}, but now fitting for the absolute rate of black hole mergers as a function of spin, rather than component spin probability distributions.
    When producing these results, we additionally infer the black hole mass and redshift distributions, as in Fig.~\ref{fig:spins-with-masses} using \textsc{Numpyro} rather than \textsc{Emcee}.
    The left-most column shows the differential merger rate $dR/dm_1\,d\chi_1\,\chi_2$ per unit primary and secondary spin magnitude, evaluated along the $\chi_1 = \chi_2$ line.
    The second column analogously shows the merger rate $dR/dm_1\,d\cos\theta_1\,d\cos\theta_2$ across the $\cos\theta_1 = \cos\theta_2$ line, and the last two columns give the rates  and $dR/dm_1\,d\chi_\mathrm{p}$ as a function of effective spins.
    In all cases, rates are evaluated at $m_1=30\,M_\odot$ and $z=0.2$.
    Even when plotting rates, we draw the same qualitative conclusions as originally identified in Fig.~\ref{fig:betaPlusDoubleGaussianTrace}: the \textsc{HighSpinPrecessing} and \textsc{MediumSpin} populations are recovered well, while the rate density of the \textsc{LowSpinAligned} population is recovered poorly.
    }
    \label{fig:spins-with-rates}
\end{figure*}

GW data are generated according to a Poisson point process, in which individual compact binaries stochastically trace an underlying \textit{rate density} $dR/d\lambda$ of mergers across the space of binary parameters $\lambda$.
When performing hierarchical inference over GW catalogs, we are formally reconstructing this rate density: measuring the ``counts'' of events occurring in different regions of parameter space.
Often, however, we are concerned only about the shape of $dR/d\lambda$, not its normalization.
In this case, it is common to instead study the normalized \textit{probability distribution} $p(\lambda)$ of source parameters.
This is achieved after the fact by fitting for the merger rate but only presenting $p(\lambda)$, or from the very outset by marginalizing over and subsequently ignoring the total rate, a procedure that gives Eq.~\eqref{eq:likelihood}.

While this procedure is usually well-behaved, there do exist instances in which the choice to present the normalized $p(\lambda)$, rather than a reconstructed rate density, can yield inadvertently misleading conclusions.
In some cases, models that successfully recover the correct rate density can appear to fail in recovering the correct probability density; see discussion in Sec.~5B of \citet{Callister:2023tgi}.
Thus, when evaluating the goodness-of-fit of a given model, the most robust results are obtained by comparing injected and recovered \textit{merger rates}, rather than injected and recovered probability densities.

Given this discussion, does the poor agreement between injected and recovered spin probability distributions signify a true modeling and inference failure?
Or is this disagreement illusory, due to our choice to compare probability distributions rather than reconstructed merger rate densities?
To check this, we repeat the hierarchical analyses of the three simulated populations but now fitting for the overall merger rate alongside the hyperparameters governing the component spin distributions.
As in Appendix~\ref{appendix:masses}, we simultaneously infer the primary mass and redshift distributions.
Figure~\ref{fig:spins-with-rates} shows our inferred merger rates as a function of spin for each injected population.
Our initial conclusions hold:
when simultaneously fitting for and presenting differential merger rates, rather than probability densities, we still find that the \textsc{HighSpinPrecessing} and \textsc{MediumSpin} populations are recovered well, but we do not successfully recover the \textsc{LowSpinAligned} population.
Hence our poor recovery of \textsc{LowSpinAligned} is a real effect, rather than a bias or misleading visualization related to our choice to marginalize over the absolute merger rate.

\subsection{Other miscellaneous checks for hierarchical inference}
\label{appendix:misc}

\begin{figure}
\centering
    \includegraphics[width=\linewidth]{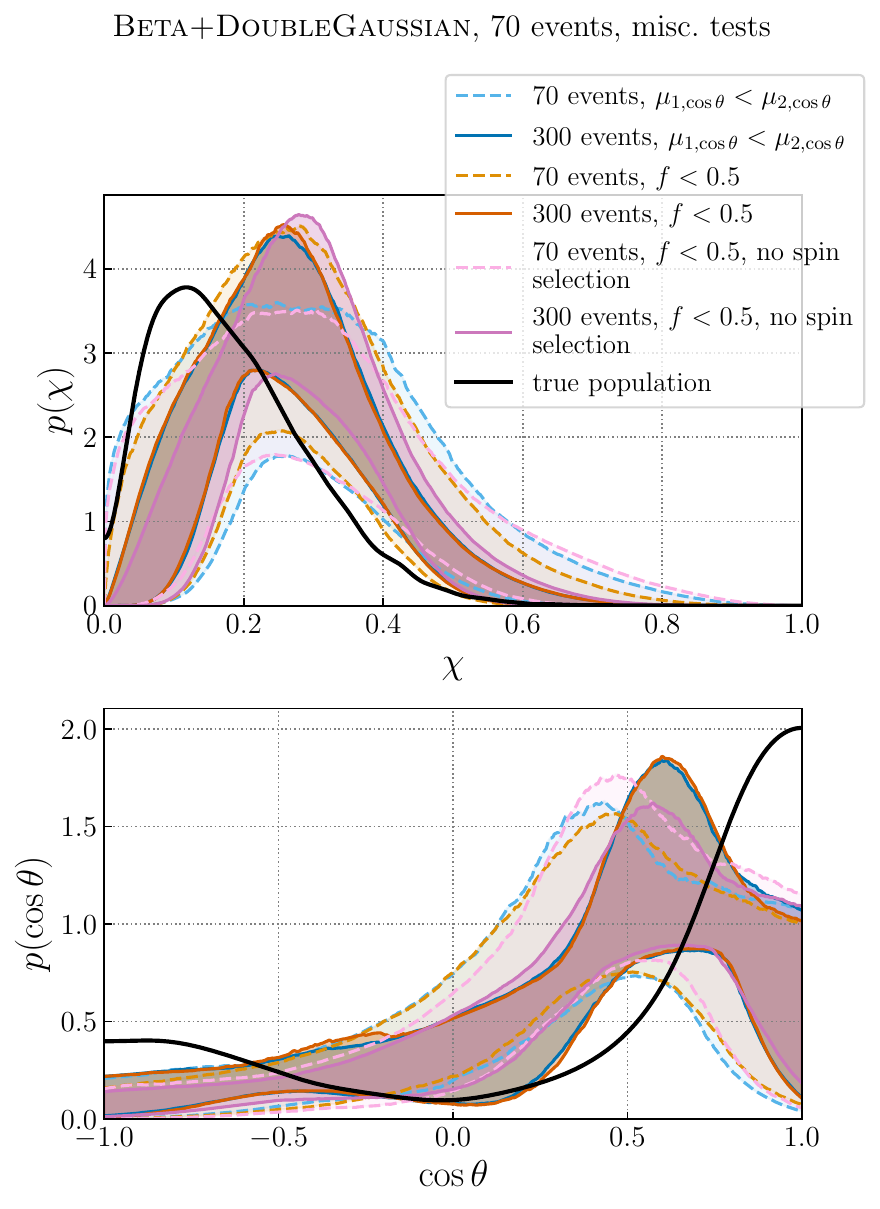}
    \caption{
     Inferred distributions (blue) for spin magnitude $\chi$ (top subplot) and cosine of the tilt angle $\cos\theta$ (bottom subplot) for 70 events from the \textsc{LowSpinAligned} population obtained with the \textsc{Beta+DoubleGaussian} population model under various conditions.
     All results from 70 (300) event catalogs are plotted with dashed (solid) lines.
     In blue and orange we compared two methods of breaking the degeneracy between the two Gaussian sub-populations in the \textsc{Beta+DoubleGaussian} tilt distribution: first, restricting the mean of distribution 1 to be smaller than that of distribution 2 ($\mu_{1,\cos\theta}<\mu_{2,\cos\theta}$; blue), and second, restricting the mixing fraction to be less than one half, effectively assigning distribution 1 to be that containing less events ($f<0.5$; orange). 
     In orange and pink, we compare results where we do (orange) versus do not (pink) include spin selection effects (both using the $f<0.5$ method of breaking degeneracy). 
     None of these variations produce population constraints improved from those shown in Fig.~\ref{fig:betaPlusDoubleGaussianTrace}; the bimodality of the tilt-angle distribution remains unrecovered.
    }
    \label{fig:misc_tests}
\end{figure}

Finally, we present results from other miscellaneous verification methods for our hierarchical inference procedure. 
First, we investigate different methods of breaking the degeneracy between the two Gaussian components in tilt-distribution portion of the \textsc{Beta+DoubleGaussian} model, see Eq.~\eqref{eq:doubleGaussian}. 
For any model defined as a mixture of multiple components, some method must be imposed to break the degeneracy these components. 
For a bimodal Gaussian, this can be done in three ways: 
\begin{enumerate}
    \item Imposing an ordering of the \textit{means} -- assign ``distribution 1" to be that with the smaller mean and ``distribution 2" to be that with the larger mean 
    \item Imposing an ordering of the \textit{widths} -- assign ``distribution 1" to be that which is narrower, and ``distribution 2" to be wider
    \item Limiting the \textit{mixing fraction} be $\leq0.5$ -- assign ``distribution 1" to be that which contains a smaller fraction of events, and ``distribution 2" to be that which contains a larger fraction.
\end{enumerate}
Sometimes one method of breaking the degeneracy converges better when used in a hierarchical inference procedure than another. We find that this is not the case in this work: different methods perform identically (within sampling error), as seen in Fig.~\ref{fig:misc_tests}. 
Using the means (mixing fraction) of the Gaussians to break the degeneracy yields the distributions plotted in blue (orange).
The results are consistent with each other. 
We opt to use the mixing fraction to break degeneracy throughout the bulk of this work because it is more computationally efficient. 

Next, to ensure there is no misspecification in the selection function for spins, see Eq.~\eqref{eq:detection-efficiency-mc}, we run hierarchical inference without any spin selection effects. 
Results are shown for the \textsc{LowSpinAligned} population pink in Fig.~\ref{fig:misc_tests} for 70 (dashed) and 300 (solid) events.
Selection effects in component spins are not strong, and thus are not expected to effect population inference significantly.\footnote{Selections effects are strong for masses and redshift, on the other hand. This can be seen when comparing the underlying and detected distributions in Fig.~\ref{fig:mass_redshift_dists}.} 
This is indeed the case, as the distributions inferred without spin selection effects are nearly identical to those inferred with them (shown in orange).
We also note that for all results shown in this work, the number of effective samples does not rail against the cut given in Eq.~\eqref{eq:Neff}.

\begin{figure*}
\centering
    \includegraphics[width=\linewidth]{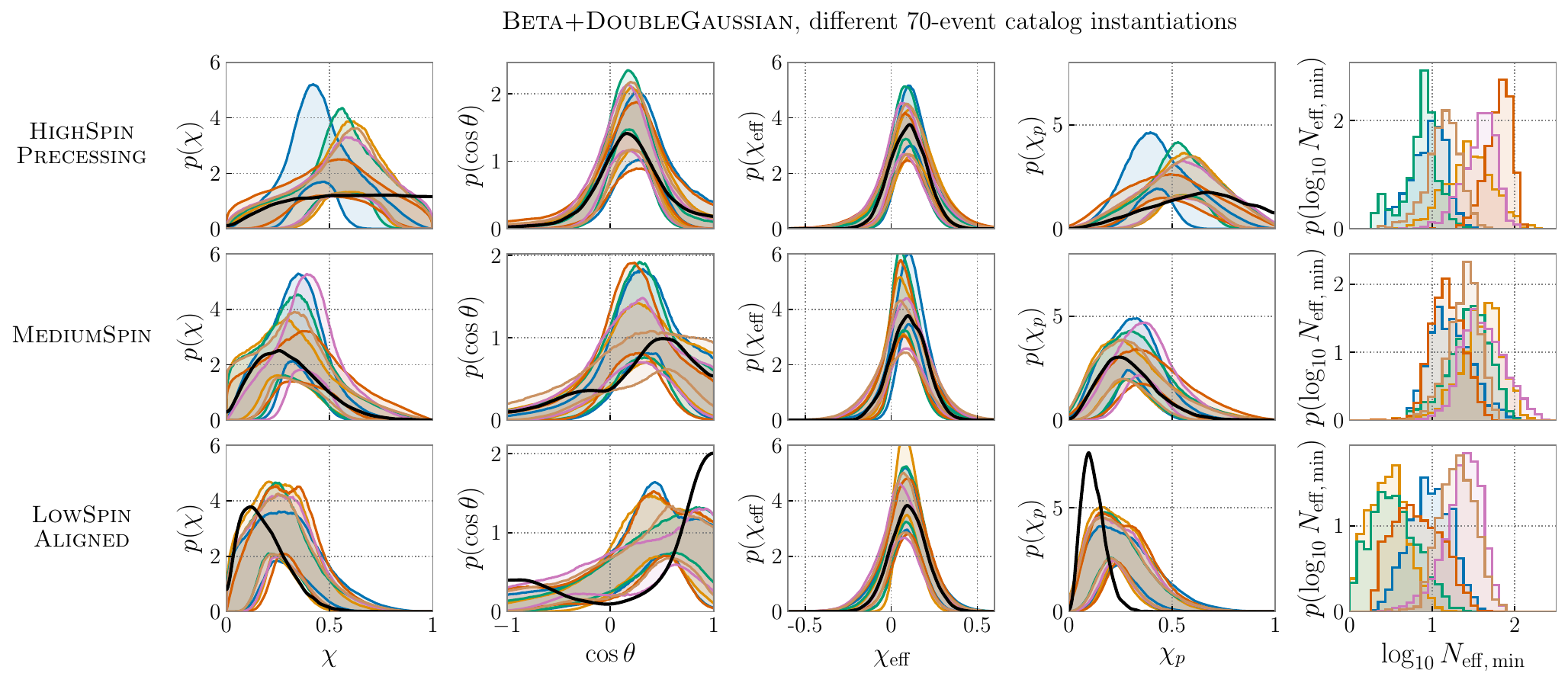}
    \caption{
     Inferred distributions obtained with the \textsc{Beta+DoubleGaussian} model for spin magnitude $\chi$, spin tilt $\cos\theta$, effective $\chieff$ spin, effective precessing spin $\chip$, for various 70-event catalog instantiations of the three simulated populations, each plotted in a different color. 
     The true, underlying populations are shown in black for comparison. 
     While there is a small amount of variation, none of the catalog instantiations yields population constraints improved from those shown in Fig.~\ref{fig:betaPlusDoubleGaussianTrace}.
     The rightmost column shows the minimum per-event $\Neff$ over the catalog used for inference for that particular analysis (See Eq.~\eqref{eqn:Neff_weights} and corresponding discussion in Appendix~\ref{appendix:inference}). 
     Within a given row, the minimum $\Neff$ does not correlate with the by-eye goodness of fit of the distributions in the first four columns. 
    }
    \label{fig:diff_catalog_instantiations}
\end{figure*}

Our final check is to conduct hierarchical inference on several different random 70-event catalog realizations from the 300 total events per population.
Just like different Gaussian noise instantiations of the data lead to variance in individual-event posteriors, so too can random catalog instantiation lead to variance in the recovered posteriors on the \textit{population} parameters. 
Some catalogs will yield a more accurately recovered population than others, just by random chance from working with finite numbers, see, e.g.~\citet{Callister:2022qwb}.
Fig.~\ref{fig:diff_catalog_instantiations} shows some expected variance in results, but nothing corresponding to the degree of mismatch between the true and inferred tilt distributions of the \textsc{LowSpinAligned} population seen in Fig.~\ref{fig:betaPlusDoubleGaussianTrace}.
We therefore conclude that we cannot attribute bias between the injected and recovered \textsc{LowSpinAligned} population to be from an ``unlucky" catalog realization. 
Additionally, each catalog instantiation leads to a different number of per-event effective samples, see Eq.~\eqref{eqn:Neff_weights} and corresponding discussion in Appendix~\ref{appendix:inference}, which we find within a given population are \textit{not} correlated to the by-eye goodness of fit, as seen in the rightmost column of Fig.~\ref{fig:diff_catalog_instantiations}. 
However, the \textsc{LowSpinAligned} population yields, on average, the lowest $\Neff$ values and is the least accurate fit.
This leads us to believe that the absence of inferred bimodality is not due to the issue of our events not having enough effective samples, although the fact that the minimum event-level $\Neff$ over catalog instantiations is small could be another source of imperfect recovery.
As part of future work, we plan to explore the uncertainty in difference in log-likelihood formulated in \citet{Talbot:2023pex} as another way to gauge whether our Monte Carlo estimations avoid bias.

\bibliography{bib}

\end{document}

%% file: macros.tex
\newcommand{\szMean}{0.10}
\newcommand{\szStd}{0.15}
\newcommand{\chieffMean}{0.10}
\newcommand{\chieffStd}{0.11}
\newcommand{\popTwoChiPeak}{0.25}
\newcommand{\popTwoChiStd}{0.20}
\newcommand{\popThreeChiPeak}{0.10}
\newcommand{\popThreeChiStd}{0.05}
\newcommand{\trueSigmaMeas}{0.48}